\documentclass[sigconf]{acmart}

\setcopyright{none}
\acmConference[Technical Report]{ACM}{ACM}{2022}
\acmYear{2022}
\usepackage{subcaption}
\makeatletter \newif\if@restonecol \makeatother  
\usepackage[linesnumbered, ruled, vlined]{algorithm2e}
\usepackage{algpseudocode}

\usepackage[bottom]{footmisc}

\usepackage{epstopdf}
\usepackage{graphics}
\DeclareGraphicsExtensions{.eps,.gz,.eps,.pdf,.png,.jpg}
\graphicspath{{./}{figures/}}

\usepackage{hyperref}

\usepackage{booktabs}
 \usepackage{makecell}
\usepackage{listings}
\usepackage{graphicx}
\usepackage{amsmath,amsfonts}
\usepackage{dsfont}
\usepackage{wrapfig}
\usepackage{enumerate}
\usepackage{paralist}
\usepackage{multirow}
\usepackage{verbatim}

\usepackage{amsthm}
\usepackage{threeparttable}
\usepackage{tabularx}

\newtheorem{myprob}{Problem}
\newtheorem{myprop}{Proposition}

\newcommand{\comments}[1]{}
\newcommand{\ignore}[1]{}
\newcommand{\eat}[1]{}

\newcommand{\sstitle}[1]{\smallskip\noindent\textbf{#1.\/}}

\DeclareMathOperator*{\argmin}{arg\,min}

\setdefaultleftmargin{1.5em}{1em}{}{}{}{}

\newcommand{\edit}[1]{#1}

\def\Snospace~{\S{}}

\makeatletter
\newcommand{\removelatexerror}{\let\@latex@error\@gobble}
\makeatother

\hypersetup{
	bookmarks=true,        
	unicode=true,          
	pdffitwindow=false,    
	pdfstartview={FitH},   
	pdftoolbar=true,       
	pdfmenubar=true,       
	colorlinks=true,      
	linkcolor={NavyBlue},       
	urlcolor={NavyBlue},        
	citecolor={NavyBlue},       
	filecolor=black,       
	pdftitle={},           
	pdfauthor={}, 
	pdfcreator={} 
}

\begin{document}





\title{Detecting Rumours with Latency Guarantees using Massive Streaming Data}


\author{
Thanh Tam Nguyen,
Thanh Trung Huynh,
Hongzhi Yin,
Matthias Weidlich,
Thanh Thi Nguyen,
Thai Son Mai,
Quoc Viet Hung Nguyen
}

\renewcommand{\shortauthors}{Thanh Tam Nguyen, et al.}





\begin{abstract}

Today's social networks continuously generate massive streams of data, which 
provide a valuable starting point for the detection of rumours as soon as they 
start to propagate. However, rumour detection faces tight latency bounds, which 
cannot be met by contemporary algorithms, given the sheer volume of 
high-velocity streaming data emitted by social networks. Hence, in this paper, 
we argue for best-effort rumour detection that detects most rumours quickly 
rather than all rumours with a high delay. To this end, we combine techniques 
for efficient, graph-based matching of rumour patterns with effective load 
shedding that discards some of the input data while minimising the loss in 
accuracy. Experiments with large-scale real-world datasets illustrate the 
robustness of our approach in terms of runtime performance and detection 
accuracy under diverse streaming conditions.

\end{abstract} 

\maketitle

\section{Introduction}

Social networks such as Twitter, Facebook, and Yelp
produce data streams at an unprecedented rate. For instance, half a billion 
tweets are generated every day~\cite{ding2019interspot}. However, social 
networks are known to  be susceptible to the spread of false information and 
rumours~\cite{vosoughi2018spread}. The inherent openness of social platforms 
enables users to produce and propagate content without authentication and 
verification, which 
has been exploited for massive rumour propagation. Reports show that 
three-quarters of online news consumers say they encounter one or more 
instances of rumours every day~\cite{ma2016detecting}.
 
The high velocity and low veracity of data streams emitted by social networks 
render rumour detection extremely challenging. Beyond the sheer volume of data 
to process, the challenges stem from the fact that a rumour detection model 
needs 
to be continuously retrained to cope with concept drift. 
Especially in case of short bursts, 
which frequently happen during the propagation of 
rumours~\cite{tam2019anomaly}, the data volume exceeds processing limits,
and the latency of rumour detection becomes unacceptably high. 
The latter is of utmost importance since rumours often cause 
devastating socio-economic damage before 
being effectively corrected~\cite{vosoughi2018spread}. The reason being that 
innocent users, without proper alerts from algorithmic models, tend to spread 
false information accidentally, thereby creating an exponentially growing 
amount of rumour representations.
Hence, rumour detection for social platforms needs to meet tight latency bounds 
in order to enable mainstream media, governments, legal agencies, and public 
organisations to react immediately on emerging rumours, false stories, and 
misinformation~\cite{srijith2017sub,farajtabar2017fake}. 

In this paper, we argue for best-effort rumour detection that aims to maximise 
the number of rumours detected under a given latency constraint. To this end, 
we propose a load shedding framework
that discards some streaming data. However, shedding might compromise 
detection accuracy, so that rumours could be missed or falsely 
detected. Thus, it is important to discard only data that has a small impact on 
rumour detection accuracy; and ideally, none at all. 
Selecting data for shedding raises several research questions, though: i) 
How to quantify the importance of data for rumour detection; ii) How to 
calculate the amount of data to shed in order to maintain a latency bound; and 
iii) How to shed data efficiently to avoid additional overhead.
\edit{However, it is not straightforward to answer these questions. 
Skipping some inputs might make the system get out of overload in a short-term but does not change the efficiency nature of rumour detection, while incurring information loss and making the detection performance worse.
Therefore, a good solution requires changing the nature of rumour detection algorithms from offline to online and designing a load shedding mechanism that takes into account these online information to retain only significant data elements. 
Moreover, these questions shall be addressed for data streams with unstable and 
fluctuating rates, which render traditional parallelization and distribution 
schemes insufficient~\cite{wang2015bigdetecting}.
}

Addressing the above questions, 
our contributions and the paper structure, following some background on 
anomaly-based rumour detection in \autoref{sec:background}, 
are summarised as follows: 

\begin{itemize}
	\item \emph{Model and Architecture for efficient streaming rumour detection:} 
	In \autoref{sec:problem}, we propose a general framework that enables 
 efficient rumour detection on high-velocity streams. At its core, this 
 framework includes a load shedding strategy to guarantee latency constraints 
 by discarding some of the input data. Using the remaining data, rumours are 
 detected online by graph-based matching of propagation patterns and an 
 anomaly scoring mechanism. 
 \item \emph{Streaming rumour detection:} In 
 	\autoref{sec:incremental_detection}, we provide techniques for 
 	pattern matching and anomaly scoring in our framework. Based on 
 	existing mechanisms for static data, we present online algorithms 
 	that enable efficient rumour detection over data streams. 
 \item \emph{Coefficient-based load shedding:} In \autoref{sec:shedding},  we 
 design a statistical model to capture the importance of social data for rumour 
 detection. It assesses the correlation between data types and their 
ordering, and the patterns hinting at rumours. 
Based thereon, we present a load shedding algorithm that balances 
the trade-off between detection accuracy and latency. 
\end{itemize}

We report on an experimental evaluation in \autoref{sec:exp}. Then, 
\autoref{sec:related} 
reviews our contributions in the light of related work, before 
\autoref{sec:conclusion} concludes the paper.

%
%

\section{Background}
\label{sec:background}

This section provides background for our work. We first discuss how the spread 
of rumours in social networks is manifested in patterns in the respective data 
(\autoref{sec:rumour_patterns}). Subsequently, we summarise how these patterns 
form the basis for anomaly-based rumour detection 
(\autoref{sec:anomaly_detection}). 

\subsection{Rumour patterns in social network data }
\label{sec:rumour_patterns}

\begin{figure}[t]
\centering
\includegraphics[width=0.65\linewidth]{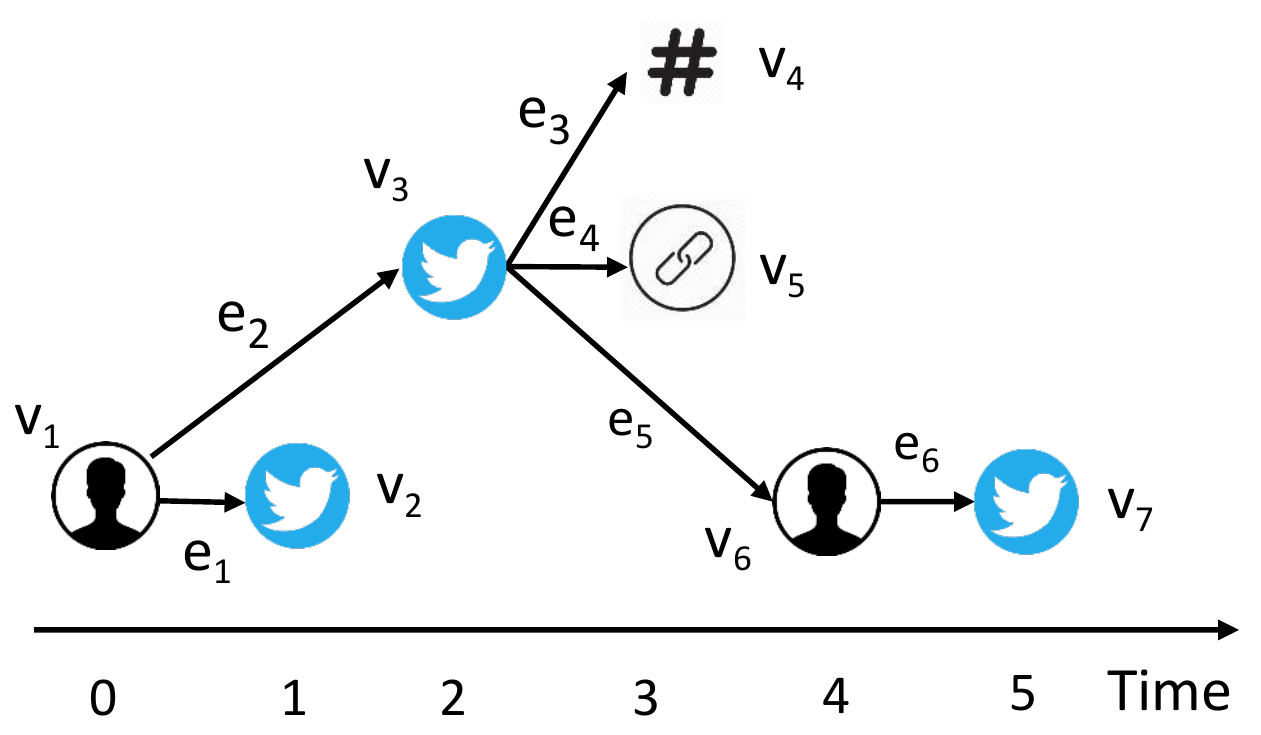}
\caption{Rumour propagation in a social network.}
\label{fig:rumour}
\end{figure}

Considering the data emitted by a social network, a rumour is a rooted 
sub-graph of a graph spanned by its entities~\cite{vosoughi2018spread}. Taking 
the case of Twitter as an example, \autoref{fig:rumour}, illustrates entities 
such as users, tweets, links, and hashtags, along with their relations. 
Assuming that the visualised subgraph denotes a rumour, the root 
entity models the user who first published the rumour as part of a tweet. The 
subgraph further indicates other related entities, such as involved 
hashtags and links, as well as the propagation of the rumour through forwarding 
and reposting by another user.
Monitoring the relations between entities, most of which 
are timestamped, reveals the propagation structures of 
rumours~\cite{tam2019anomaly}.

While different characteristics of social networks are linked to rumours, 
graph-based patterns of the propagation structure are known to be most relevant 
to detect rumours~\cite{vosoughi2018spread,drimux2016}. Using data on the 
relations between the network's entities, 
rumours can be detected based on patterns, so-called \emph{rumour 
patterns}, dozens of which have been identified for diverse 
domains~\cite{wang2017early,ma2016detecting}. 

\begin{figure}[!h]
\centering
\includegraphics[width=0.65\linewidth]{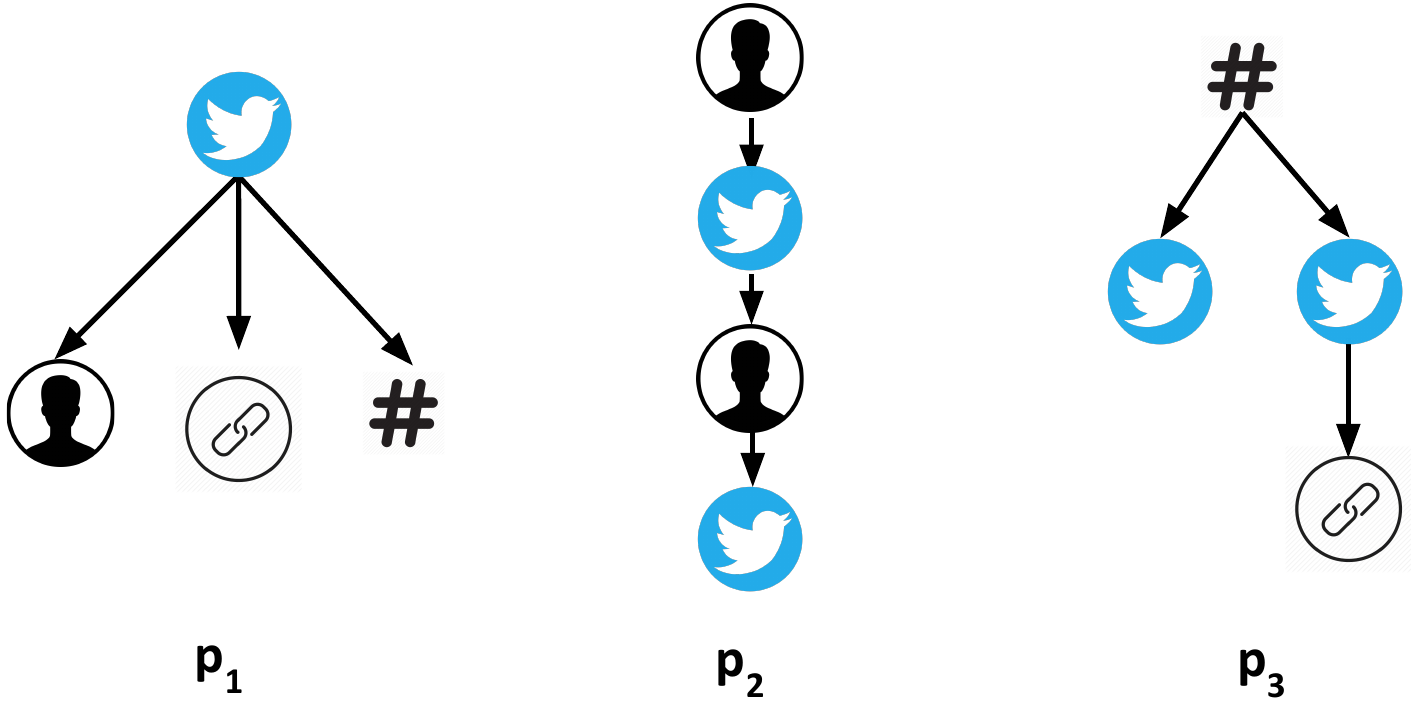}
\caption{Typical rumour patterns in social networks.}
\label{fig:pattern}
\end{figure}

\begin{example}
\autoref{fig:pattern} illustrates rumour patterns for the case of 
Twitter. In pattern $p_1$, a tweet that initiates a rumour 
contains a link, a hashtag, and a mention of a famous person. Pattern $p_2$ 
reflects a cascade of retweets by users. Pattern $p_3$ captures 
that a rumour often originates from a hot topic, which is captured by a hashtag 
that is used frequently. 
\end{example}

Contemporary approaches to rumour detection use the matches of such rumour 
patterns as part of a learning process. 
That is, the matches are used as indicators of potential rumours rather 
than immediate characterizations of rumours. They are fed into a rumour 
detector, which then decides whether a pattern match shall be considered a 
rumour, or not. Note that a match also include node and edge features of the 
subgraph, which help a rumour detector to separate rumours from other social 
events. While most existing proposals for rumour detection consider the 
subgraph structure, some techniques also consider only the set of entities of a 
subgraph and their associated features to classify 
rumours~\cite{castillo2011information,yang2012automatic,zhao2015enquiring}. 
Nevertheless, matching of rumour patterns is required to identify the sets of 
entities to feed into a rumour detector.
	
The exact definition of rumour patterns induces a trade-off: The more complex 
the patterns, the better 
they are correlated to rumours, separating them from other social 
phenomena~\cite{tam2019anomaly}. Yet, faced with high volumes of data as 
emitted by social networks, matching complex patterns potentially exceeds 
processing limits. Against this background, several collections of patterns 
that proved useful have been presented in recent years, 
see~\cite{tam2019anomaly,wang2015bigdetecting}. 

In our work, we largely abstract from the details of existing proposals 
for rumour detection. Rather, we focus on the efficiency of the first step of 
rumour detection, which is the matching of rumour patterns under latency 
bounds. As such, we rely on existing collections of rumour 
patterns rather than designing new ones, while our contributions are 
independent of the exact choice of a pattern collection.
\edit{Moreover, we change the nature of rumour detection from offline to online by proposing streaming pattern matching via a pattern index and streaming anomaly computation via sketch structures.}

In the remainder, we use Twitter as an example of a social network. Note though 
that our approach adopts a generic graph-based model that is applicable also to 
other social networks~\cite{shi2017survey}.

\subsection{Anomaly-based rumour detection}
\label{sec:anomaly_detection}

While rumour patterns capture common propagation structures of rumours, not all 
of their matches denote actual rumours~\cite{kwon2017rumor}. Rather, 
state-of-the-art approaches use these matches to extract features of entities 
and relations, and learn a 
model to classify matches as rumours or other types of social 
phenomena~\cite{zubiaga2018detection,vosoughi2018spread}. 
Specifically, this classification is based on inconsistencies in the 
propagation structure described by pattern matches. In a first step, this 
involves the identification local anomalies related to the features of the 
entities and relations in the pattern matches. Since such local anomalies, 
typically, do not represent reliable signals, a second step is the 
identification of {global anomalies} that comprise connected entities for which 
local anomalies have been observed. Below, we exemplify this general approach. 

\begin{figure}[t]
  \centering
  \includegraphics[width=0.85\linewidth]{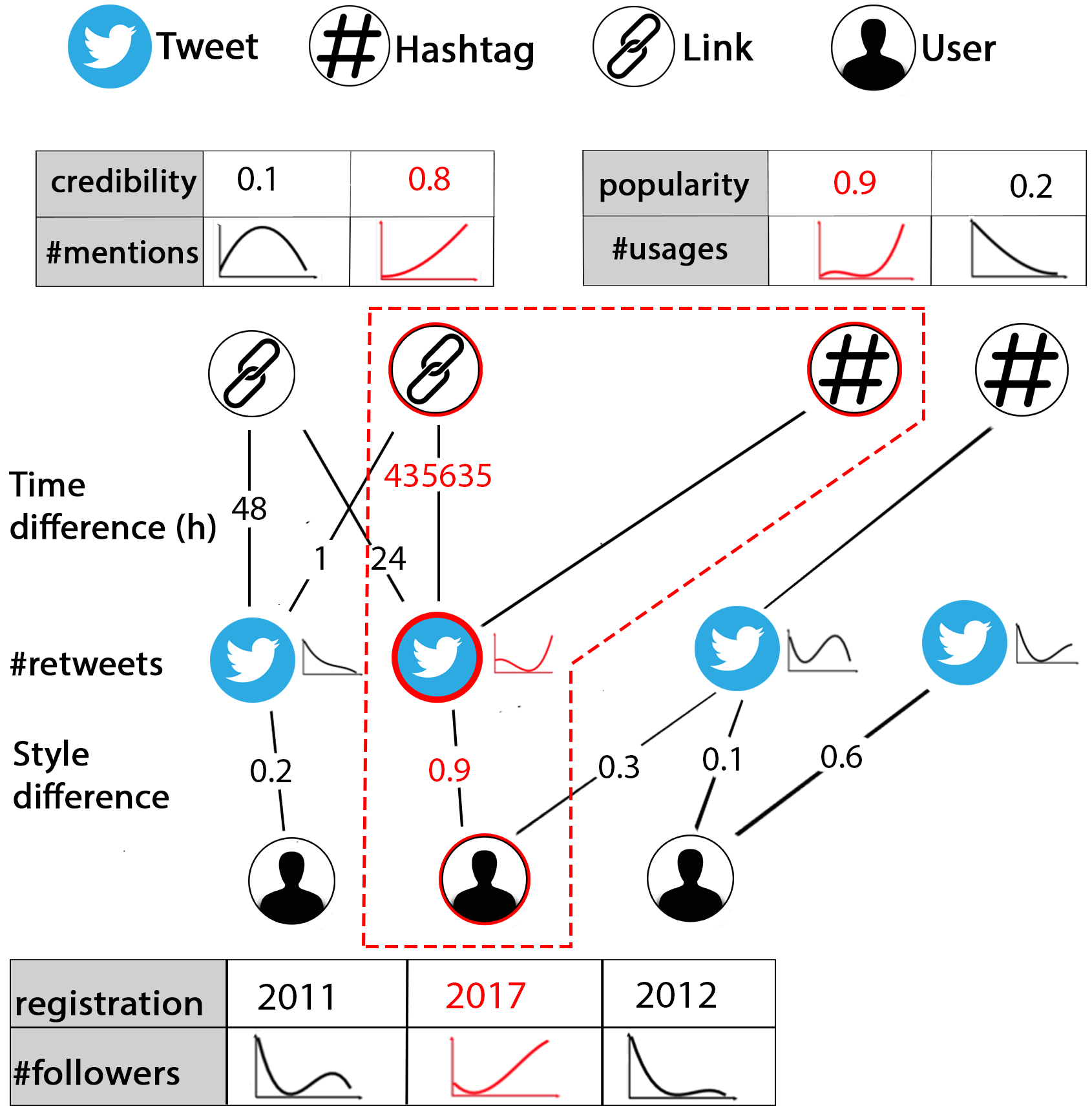}
  \caption{Anomalies in the propagation structure, see~\cite{tam2019anomaly}.}
  \label{fig:entity}
\end{figure}

\begin{example}
Consider the Twitter social network in \autoref{fig:entity}. Here, entities 
have features, e.g., a user
has a registration date and a number of followers. Also, relations are 
annotated with attributes, e.g., the tweet-article relation is annotated with 
the difference between the publication dates of the tweet and the article.
\autoref{fig:entity} illustrates how rumours are manifested in 
local anomalies: The highlighted user has a registration date that is 
significantly newer than those of related users and the number 
of retweets of the highlighted tweet is suddenly very high.
Using these local anomalies, a global anomaly is identified: The highlighted 
subgraph in the social network comprises a user, a tweet, a hashtag, and a 
linked article that all show local anomalies. In addition, some of the 
relations in the subgraph provide evidence for a rumour, such as the 
time difference between the highlighted tweet and linked article. 
\end{example}
  
In the remainder of this section, we illustrate in more detail the features 
used for rumour detection along with the respective scoring mechanisms.
   

\subsubsection{Features to identify rumours}
\label{sec:features}

While the definition of meaningful features is specific to the domain of the 
considered social network, we discuss some of the features defined 
in~\cite{tam2019anomaly} for the case of Twitter as illustrative examples. 
Considering the entities, for instance, the registration age and a credibility 
score are used to identify rumours. Also, sudden changes in the number of 
followers of a user or the frequency with which a user publishes status updates 
potentially indicate anomalous behaviour. For articles linked in tweets, the 
linguistic style as well as the number of {mentions} over time are 
incorporated. 
%
%
%
Turning to the relations in a social network, for instance, differences in the 
time and location of a tweet and a linked article serve as features. Another 
example is the relation between users and hashtags, for which the fact that the 
user has not used the hashtag before represents a feature. 

The combination of a large number of diverse features ensures a certain 
robustness of rumour detection.

\subsubsection{Local and global anomaly computation}
\label{sec:local_global}

Using sets of features as discussed above, rumour detection first identifies 
local anomalies related to individual entities and relations, before 
aggregating them to identify global anomalies, i.e., in terms of a subgraph in 
the social network~\cite{tam2019anomaly}.

\sstitle{Element-level scoring}
An anomaly score is derived based on the differences between the
current and past values of a feature vector assigned to an entity or a 
relation. To this end, an anomaly score is first derived per feature. Using 
historic data, this score is computed as a non-parametric 
statistical measure, an empirical p-value, that quantifies the number of past 
feature values that differ from the current observation. These anomaly scores 
per feature are then aggregated for each entity and relation, respectively. 
It was argued that such aggregation shall be guided by the minimum values over 
all features~\cite{tam2019anomaly}. This way, false positives that stem from a 
few features that indicate anomalies, whereas other features do not, shall be 
avoided. 

\sstitle{Subgraph scoring}
Using the local anomaly scores per entity and relation, a global anomaly score 
is computed for connected subgraphs to identify those that represent a rumour. 
To this end, scan statistics~\cite{kulldorff1997spatial} that assess the 
statistical significance of a subgraph being anomalous can be leveraged. In 
essence, it considers all elements (entities and relations) with anomaly scores 
that are significant at a given confidence level. To capture the propagation of 
rumours, the employed detection mechanism is commonly designed such that 
elements that have insignificant scores may be added to the subgraph as long as 
they are connected with a sufficiently large number of elements with 
significant scores.

\section{Efficient Streaming Rumour Detection}
\label{sec:problem}

This section introduces a general framework for efficient rumour detection on 
high-velocity streams. 
\edit{We argue that a good solution requires changing the nature of rumour detection algorithms from offline to online and designing a load shedding mechanism that takes into account these online information to retain only significant data elements.}
In \autoref{sec:architecture}, we first present a 
respective architecture. Subsequently, \autoref{sec:model} defines a formal 
model that captures the most important notions needed to the instantiate this 
architecture in the remainder of the paper.

\subsection{Architecture}
\label{sec:architecture}

\begin{figure}[t]
  \centering
  \includegraphics[width=.95\linewidth]{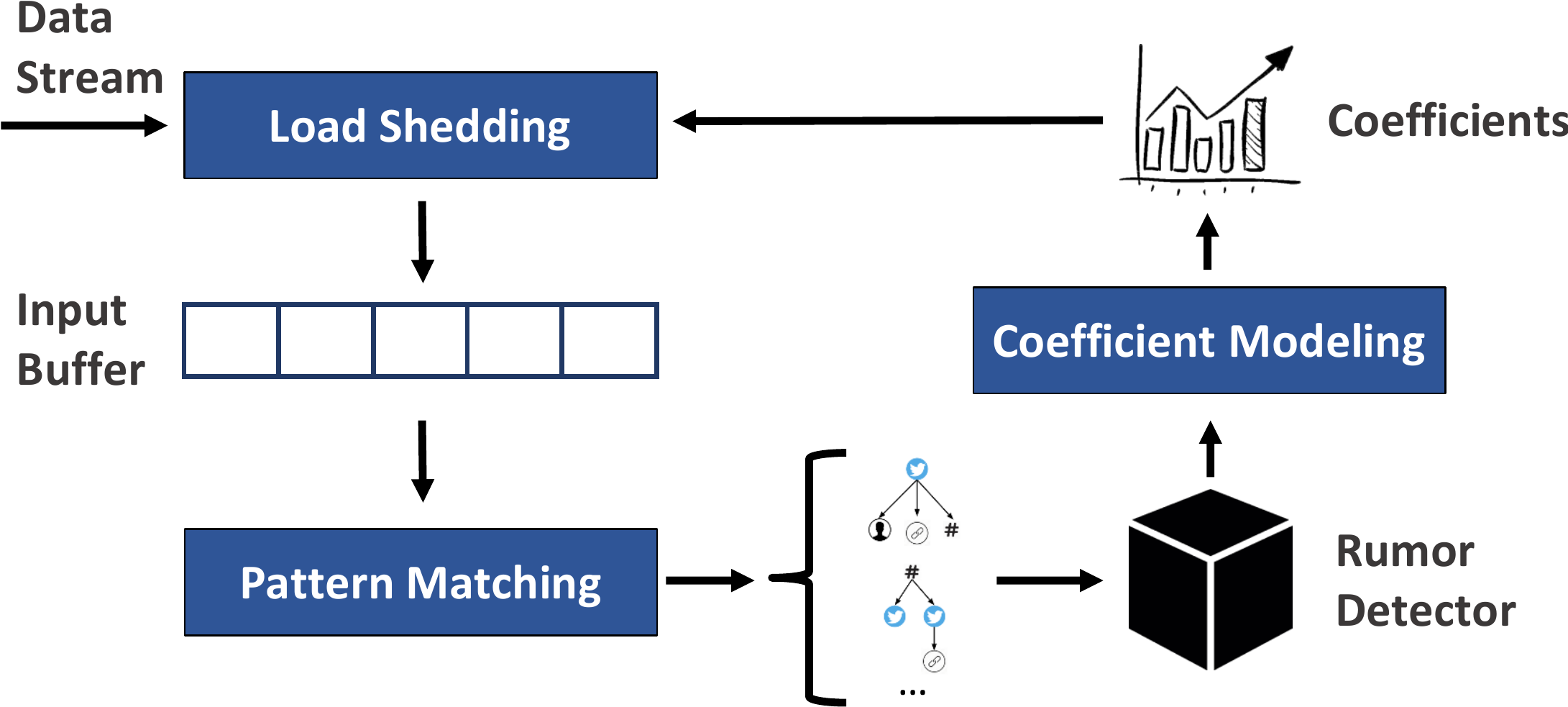}
  \caption{Architecture for streaming rumour detection.}
  \label{fig:framework}
\end{figure}

In \autoref{fig:framework}, we outline a general architecture for streaming 
rumour detection. Ignoring load shedding for the time being, streaming data 
emitted by a social network are pushed to an input buffer. They are then 
processed by \emph{pattern matching}, which is based on 
a predefined collection of patterns, as exemplified in the previous section. 
The matches are used by a \emph{rumour detector} to identify rumours. Here, we 
abstract from the internals of the detector, such as the exact set of 
employed features. Instead, we focus on the algorithms used to 
compute anomaly scores online based on some set of features and the matches of rumour 
patterns.


This general approach is extended with functionality for \emph{load shedding} 
in order to achieve efficient processing of the stream. Aiming to meet tight 
latency bounds, the respective component needs to assess whether the rumour 
detector is overloaded, which is operationalised by checking the input buffer 
size periodically. 
If an overload situation is detected, load shedding discards some input data 
before it is considered for pattern matching, thereby preventing that a latency 
threshold is violated. 

\emph{Coefficient modelling}, in turn, uses the results of the rumour detector 
to guide load shedding. Here, the goal is to minimises the loss 
in detection accuracy implied by the discarded data. To this end, a model of  
coefficients is maintained that captures the importance of particular types of 
data in the input buffer for rumour detection. 

In sum, our approach is to monitor the processing rate 
(via waiting time of data in the buffer) and the utility of data (via 
statistics) to shed some data elements in the buffer when the system is 
overloaded (the latency threshold is violated). Intuitively, if we shed 
important data elements (high utility), it would harm the rumour detection 
accuracy as the detector will not have sufficient information. However, if less 
important data is kept, the system may become overloaded. This issue is 
addressed in our approach with statistical coefficients to capture the utility 
in the long run and to estimate the utility of any new data element.

Both, load shedding and maintenance of coefficients are time-critical 
tasks, as they shall ensure efficient processing in an overload situation. 
Hence, they must not induce a notable overhead, since this would thwart any 
performance benefit resulting from the discarded data.


\subsection{Model}
\label{sec:model}

We capture the setting of rumour detection over streaming data with the 
following notions and notations. An overview is provided in 
\autoref{tab:notation}.

\begin{table*}
	\caption{Overview of notions and notations.}
	\label{tab:notation}
	\centering
	\footnotesize
	\begin{tabular}{l@{\hspace{.6em}}p{13.3cm}}
		\toprule
		Notation & Meaning\\
		\midrule
		$\mathcal{M}$ & Set of modalities\\
		$\mathcal{U}$ & Set of entities\\
		$G^{(t)} = (V^{(t)}, E^{(t)}, M^{(t)})$ & Social graph at time $t$: 
		Entities $V^{(t)}$, relations $E^{(t)} \subseteq 
		V^{(t)}\times V^{(t)}$, and modalities $M^{(t)}: V^{(t)} \rightarrow 
		\mathcal{M}$\\
		$\mathcal{S} = \{{s}_1, {s}_2, \ldots 
		\}$ & Stream of 
		timestamped elements ${s}_i=(u,u',m,m',t)$ with $u,u'\in \mathcal{U}$ 
		and 
		$m,m'\in \mathcal{M}$\\
		$\mathcal{S}(t)$ & Prefix of the stream up to time $t$\\		
		$w=\{s_{1},\ldots,s_{|w|}\}$ & Window of $|w|$ stream elements, $s_i 
		\in \mathcal{S}$, $1\leq i\leq |w|$\\		
		$p=(V_p,E_p,M_p)$ & Rumour pattern: Entity variables $V_p$, 
		relation variables $E_p \subseteq V_p\times V_p$, and 
		modalities $M_p: V_p \rightarrow \mathcal{M}$\\
		$P=\{p_1, \ldots, p_q\}$ & Set of rumour patterns\\
		$\lambda: V_p \rightarrow V^{(t)}$ & 
		Bijection capturing a match of pattern $p=(V_p,E_p,M_p)$ over $G^{(t)} 
		= 
		(V^{(t)}, E^{(t)}, M^{(t)})$, such that 
		$(v,v')\in E_p \Leftrightarrow (\lambda(v),\lambda(v'))\in E^{(t)}$ and 
		$\forall \ v\in V_p: M_p(v)= M^{(t)}(\lambda(v))$\\
		$\Gamma$ & Rumour detector\\
		$f(\Gamma,P,\mathcal{S}(t))$ & Detection coefficient: \# rumours 
		detected by $\Gamma$ over stream prefix $\mathcal{S}(t)$ based on 
		patterns $P$\\
		$\rho: \mathcal{S} \rightarrow \{0,1\}$ & Load shedding function: 
		Indicates which stream elements to discard ($0$) or process ($1$)\\
		$\mathcal{S}'(t) = \rho(\mathcal{S}(t))$  & Stream prefix after load 
		shedding\\
		\bottomrule
	\end{tabular}
\end{table*}

\sstitle{Multi-modal social graph}
Let $\mathcal{M}$ be a set of modalities and 
$\mathcal{U}$ be a set of entities.
Then, a multi-modal social graph at a specific time point 
$t$ is an annotated graph $G^{(t)} = (V^{(t)}, 
E^{(t)}, M^{(t)})$, where 
vertices $V^{(t)}\subseteq \mathcal{U}$ denote a set of entities, edges 
$E^{(t)} 
\subseteq V^{(t)}\times V^{(t)}$ denote a set of 
relations, and $M^{(t)}: V^{(t)} \rightarrow 
\mathcal{M}$ is a function that assigns a modality to each entity. Note that 
the latter induces a second-order modality for each relation, i.e., the pair of 
modalities of the source and target entities. 
The notion of a social graph enables us to address rumour detection in a 
very generic setting~\cite{hao2018social}.

\begin{example}
  We construct a multi-modal social graph from the example given in 
  \autoref{fig:rumour} as 
  follows. The example comprises four modalities, which are 
  given as $\mathcal{M}=\{user, 
  tweet, hashtag, 
  link\}$. Then, at time point $5$, the set of entities can be denoted by 
  $V^{(5)}=\{v_1,v_2,\ldots,v_7\}$, where, for instance, $M^{(5)}(v_1)=user$ 
  and  $M^{(5)}(v_2)=tweet$. The set of relations is $E^{(5)}=\{e_1, 
  \ldots, e_6\}$. Moreover, as an example, the second-order modality of 
  relation $e_1$ is $(user,tweet)$.
\end{example}

\sstitle{Data stream}
We model the stream of data emitted by a social network by an infinite set of 
timestamped stream elements $\mathcal{S} = \{{s}_1, {s}_2, \ldots 
\}$. Each such element ${s}_i=(u,u',m,m',t)$ contains two entities, 
$u,u'\in \mathcal{U}$, of modalities $m,m'\in \mathcal{M}$, respectively, while 
$t$ is an occurrence timestamp. The occurrence timestamps of elements in a 
stream increase monotonically, which is denoted by their subscript. With the 
arrival of ${s}_i$, $u$ and $u'$ 
become vertices of $G$, if they are not already part of the graph, while 
$(u,u')$ becomes an edge. This way, both, new entities and new relations 
between them are modelled as part of the stream.
By $\mathcal{S}(t) = \{ (u,u',m,m',t') \in \mathcal{S} \mid  t'\leq t \}$, we 
denote the stream prefix up to time $t$.

The data emitted by social networks is typically partitioned into windows 
(e.g., as extracted from the Twitter Streaming API). Here, we consider windows 
of a fixed size (count-based windows), even though our 
approach can also be applied for time-based windows, predicate-based windows, 
and hybrid windows~\cite{slo2019espice}. That is, a window is a set 
$w=\{s_{1},\ldots,s_{|w|}\}$ of $|w|$ elements of the stream, $s_i \in 
\mathcal{S}$, $1\leq i\leq |w|$. 
For the sake of brevity, we consider windows to be disjoint. However, we note 
that overlapping windows could be handled as well once a hashing mechanism 
would be in place to avoid duplicates.

\sstitle{Rumour patterns}
As detailed above, contemporary approaches to rumour detection
exploit collections of rumour patterns. 
With $\mathcal{M}$ as a set of modalities, we formalise a rumour 
pattern as an annotated graph $p=(V_p,E_p,M_p)$, where $V_p$ is a set of entity 
variables, edges $E_p\subseteq V_p\times V_p$ denote relation variables, and 
$M_p: V_p \rightarrow \mathcal{M}$ is a function that assigns a modality to 
each entity variable. A rumour detection technique processes the relations 
given in a data stream and 
matches a set of such patterns $P=\{p_1, \ldots, p_q\}$.


Let $G^{(t)} = (V^{(t)}, E^{(t)}, M^{(t)})$ be a multi-modal social graph at 
time point $t$. Then, a match of a pattern 
$p=(V_p,E_p,M_p)$ is given as an isomorphic subgraph of $G^{(t)}$. That is, a 
match is characterised by a bijection $\lambda: V_p \rightarrow V^{(t)}$, such 
that $(v,v')\in E_p \Leftrightarrow (\lambda(v),\lambda(v'))\in E^{(t)}$ and 
for 
all $v\in V_p$ it holds that $M_p(v)= M^{(t)}(\lambda(v))$.

\begin{example}
	\label{ex:pattern_match}
	Consider rumour pattern $p_2$ from \autoref{fig:pattern}, which models a 
	cascade of retweets. We formalise this pattern as $p_2 = 
	(V_{p_2},E_{p_2},M_{p_2})$, where the entity variables are given as
	$V_{p_2}=\{v'_1,\ldots,v'_4\}$, edges are defined as 
	$E_{p_2}=\{(v_1,v_2),(v_2,v_3),(v_3,v_4)\}$, and the modalities are given 
	as $M_{p_2}(v_1)=M_{p_2}(v_3)=user$ and 
	$M_{p_2}(v_2)=M_{p_2}(v_4)=tweet$.	
	Evaluating this pattern over the multi-modal social graph from 
	\autoref{fig:rumour}, we derive a match based on the following bijection, 
	$\lambda(v_1')=v_1$, $\lambda(v_2')=v_3$, $\lambda(v_3')=v_6$, and 
	$\lambda(v_4')=v_7$.
\end{example}

\sstitle{Detection coefficient}
Following the architecture outlined in \autoref{fig:framework}, a rumour 
detector derives the matches of 
rumour patterns and, based thereon, distinguishes rumours from other phenomena 
commonly observed in social networks~\cite{xing2016hashtag}. Let $\Gamma$ be 
the rumour detector, $\mathcal{S}(t)$ be a stream prefix, and $P$ a set of 
rumour 
patterns. Then, we denote by $f(\Gamma,P,\mathcal{S}(t))$ the \emph{detection 
coefficient}, which is defined as the number of rumours detected by the rumour 
detector $\Gamma$ over stream prefix $\mathcal{S}(t)$ based on the set of 
rumour patterns $P$.

\sstitle{Load shedding}
Finally, we clarify, in abstract terms, how to incorporate load shedding in 
our model. It is captured by an indicator function $\rho: \mathcal{S} 
\rightarrow \{1,0\}$ over the stream elements, specifying whether they are 
discarded ($0$) or processed ($1$). 
Applying such a function to the data stream yields a projection of the original 
stream, i.e., all elements for which $\rho$ signals that they shall be 
discarded are removed from the stream. Overloading notation, we denote 
this application by $\mathcal{S}'(t) = \rho(\mathcal{S}(t))$ for a stream 
prefix $\mathcal{S}(t)$ and by
$w' = \rho(w)$ for a window over the stream $\mathcal{S}$,
 where $\mathcal{S}'(t) 
\subseteq \mathcal{S}(t)$ and $w' 
\subseteq w$, respectively.

\section{Streaming Rumour Detection}
\label{sec:incremental_detection}

\edit{Loading shedding would make the processing system faster in short-term but do not change the efficiency nature of the rumour detection algorithm. Therefore, a good solution requires changing the nature of rumour detection algorithms from offline to online.}
In order to realise our architecture for streaming rumour detection, see 
\autoref{fig:framework}, this section clarifies how the detection can be done 
online. In particular, we introduce techniques for streaming pattern 
matching (\autoref{sec:incremental_matching}) and streaming anomaly 
computation as conducted by the rumour detector 
(\autoref{sec:incremental_anomaly}), and combine them accordingly 
(\autoref{sec:incremental_integration}).

\subsection{Streaming Pattern Matching}
\label{sec:incremental_matching}

\sstitle{Initial social graph}
The evolution of a social network is captured by a stream, an infinite set of 
timestamped relations, as introduced in \autoref{sec:model}. However, rumour 
detection typically does not start from scratch, i.e., without any information 
about the considered social network. Rather, rumour detection typically  
incorporates a snapshot of the 
network at a certain point in time. This model then serves as the basis for the 
construction of the features as well as the local and global anomaly scoring, 
as detailed in \autoref{sec:anomaly_detection}.

To capture this setting in our model, an initial social graph is derived 
from historic data of the social network.  
This data is analysed syntactically to extract different types of entities 
(e.g., the posts, links, hashtags, and users in case of Twitter) and relations 
(e.g., retweets, mentions, links, or hashtags). 
As such, the data stream that is actually processed for rumour detection 
captures the evolution of this initial social graph. 

\sstitle{Pattern index (P-index)}
To enable efficient rumour detection, we introduce a data structure, coined 
P-index. It captures for each specific entity, how many entities of a 
particular modality are connected to it. Put differently, it captures the 
in-degree and the out-degree of an entity in the social graph $G^{(t)} = (V^{(t)}, 
E^{(t)}, M^{(t)})$, grouped per 
modality. We model the P-index as two functions $I^{(t)}_{in}:V^{(t)}\times 
\mathcal{M}\rightarrow \mathbb{N}_0$ and $I^{(t)}_{out}:V^{(t)}\times 
\mathcal{M}\rightarrow \mathbb{N}_0$, so that, at time point $t$, $I^{(t)}_{in}(v,m)=k$ 
captures that there are $k$ entities of modality $m$ connected to entity $v$, while 
$I^{(t)}_{out}(v,m)=k$ means that $k$ 
entities of modality $m$ are connected from it. 
By maintaining these structural dependencies 
between entities, our index supports streaming pattern 
discovery as it tracks partial matches of each pattern, as discussed next.

\begin{example}
\label{ex:index}
	Let us illustrate the use of the P-index with the example network in 
	\autoref{fig:rumour}. At time $t=3$, the P-index of entity $v_3$ contains 
	the following entries:
	\begin{align*}
	&I^{(t)}_{in}(v_3,user)=1,& \quad &I^{(t)}_{in}(v_3,tweet)=0,\\
	&I^{(t)}_{in}(v_3,hashtag)=0,& \quad &I^{(t)}_{in}(v_3,link)=0,\\ 
	&I^{(t)}_{out}(v_3,user)=0,& \quad &I^{(t)}_{out}(v_3,tweet)=0,\\ 
	&I^{(t)}_{out}(v_3,hashtag)=1,& \quad &I^{(t)}_{out}(v_3,link)=1. 
	\end{align*}
	Using this 
	index, we know that $v_3$ does not yet belong to a match of $p_1$ in 
	\autoref{fig:pattern}, as this would require:
	\begin{align*}
		&I^{(t)}_{out}(v_3,user)=1,\\
		&I^{(t)}_{out}(v_3,hashtag)=1,\\
		&I^{(t)}_{out}(v_3,link)=1.
	\end{align*}
	Therefore, 
	although we do not save all 
	partial matches for an entity or relation, we may conclude that it is not 
	yet part of a match. Once the necessary condition for a match is satisfied, 
	we only need to conduct an additional breadth-first search to actually 
	check for a match, as will be clarified next. 
\end{example}

\sstitle{Matching algorithm}
Given a data stream and a set of 
rumour patterns, the procedure to derive the pattern matches is formalised in 
\autoref{alg:match}. 
As discussed in \autoref{sec:rumour_patterns}, the set of rumour patterns 
is pre-defined and taken from existing work~\cite{tam2019anomaly,wang2017early}.
First, the social graph is initialised as discussed above 
(\autoref{line:initg}). Then, for each element $s$ of the stream arriving at time $t$, 
the entities, the relation, and the modalities are added to the social $G^{(t)}$ 
(\autoref{line:addIfNew1} to \autoref{line:addIfNew2}).
The P-index is updated for both entities, increasing the in- and 
out-degrees for the respective modalities (\autoref{line:updateIndex1} and 
\autoref{line:updateIndex2}).
Next, we iterate over the set of patterns $P$ to identify the matches. 
To this end, we first assess whether the edge created in the social graph by the current 
stream element $s$ matches one of the relation variables of pattern $p$, which is 
a necessary condition for further matching (\autoref{line:subgraphif}). 
If this condition is satisfied, function \emph{BFSmatch} returns any new matches, or the 
empty set if there are none (\autoref{line:getMatch}).
Since the P-index maintains all 
previous connection-related information of the entities in the social graph, the index 
is sufficient to check whether there exists a match. That is, using a 
guided breadth-first search (BFS)~\cite{hu2019direction} from $(u,u')$ using the P-index and 
pattern $p$ as pivot, potential matches are derived efficiently.
This is due to the fact that the BFS can stop early, if the P-index of an 
entity does not have sufficient in- and out-degrees for a given 
pattern to match. Moreover, the search also stops when a match is found. 
Therefore, the expansion is never larger 
than the pattern itself.

\begin{example}
Let us continue with \autoref{ex:index} and consider the time that 
	$e_5$ is inserted into the social graph as an element of data stream. Then, 
	the P-index of entity $v_3$ will contain the following non-zero entries 
	(the entries with a value of zero are not shown for brevity):
	\begin{align*}
		&I^{(t)}_{in}(v_3,user)=1,\\
		&I^{(t)}_{out}(v_3,user)=1,\\
		&I^{(t)}_{out}(v_3,hashtag)=1,\\
		&I^{(t)}_{out}(v_3,link)=1.
	\end{align*}
	The P-index of entity $v_6$ will be $I^{(t)}_{in}(v_6,tweet)=1$ (again, 
	zero-value indexes are not listed for brevity). As can be seen in 
	\autoref{fig:pattern}, $e_5$ is a part of pattern $p_1$ (the relation from 
	a tweet to a user). Our algorithm will perform a BFS starting from $e_5$. 
	First, it will verify the P-index of $v_3$, i.e., it will check whether it 
	is already connected to a link and a hashtag, which is true at that time. 
	Then, the BFS will consider $e_3$ and $e_4$, before it stops, 
	as the pattern $p_1$ does not expand beyond that.
	
	Similarly, according to \autoref{fig:pattern}, $e_5$ is also a part of 
	pattern $p_2$. However, since $v_6$ is not connected to another tweet yet, 
	as indicated by the P-index, the BFS stops early and an empty match is 
	returned.
\end{example}

\begin{figure}[t]
  \begin{minipage}[t]{1.0\linewidth}
    \removelatexerror
    \begin{algorithm}[H]
      \footnotesize
      \caption{Matching of rumour patterns.}
      \label{alg:match}
      \SetKwInOut{Input}{input}
      \SetKwInOut{Output}{output}

      \Input{
        A data stream $\mathcal{S}$;\\
        \ a set of rumour patterns $P$.
      }
      \Output{
		Matches $\Lambda$ of query patterns.
      }

      \BlankLine
      Initialise social graph $G^{(0)}=(V^{(0)},E^{(0)},M^{(0)})$\label{line:initg}\;
      \BlankLine
      $I^{(0)}_{in},I^{(0)}_{out}\gets \emptyset$\tcp*{Initialise empty index}
      \BlankLine
      $\Lambda\gets \emptyset$\tcp*{Initialise empty set of matches}
      \BlankLine
      $t' \gets 0$\tcp*{Initialise current time}
      \BlankLine
      \For{${s} = (u,u',m,m',t) \in \mathcal{S}$ \label{line:for}}{

		$V^{(t)}\gets V^{(t')} \cup \{u,u'\}$\label{line:addIfNew1}\tcp*{Add entities}
		$E^{(t)} \gets E^{(t')} \cup \{(u,u')\}$\tcp*{Add relation} 
		\tcp{Add modalities of new entities}
		$M^{(t)}(v) \gets 
		\begin{cases}
		m & \text{if } v=u\\
		m' & \text{if } v=u'\\
		M^{(t')}(v) & \text{otherwise}\\
		\end{cases}$\;
		$G^{(t)}\gets (V^{(t)},E^{(t)},M^{(t)})$\label{line:addIfNew2}\;
		
      \BlankLine
		\tcp{Update pattern index}
		$I^{(t)}_{out}(v,o) \gets 
		\begin{cases}
		I^{(t')}_{out}(v,o) + 1 & \text{if } v=u \land o=m' \\
		I^{(t')}_{out}(v,o) & \text{otherwise}\\
		\end{cases}$\label{line:updateIndex1}\;  
		$I^{(t)}_{in}(v,o) \gets 
		\begin{cases}
		I^{(t')}_{in}(v,o) + 1 & \text{if } v=u' \land o=m \\
		I^{(t')}_{in}(v,o) & \text{otherwise}\\
		\end{cases}$\label{line:updateIndex2}\;  
      \BlankLine
      \tcp{For each pattern}
   		\For{$p=(V_p,E_p, M_p) \in P$}{
			\tcp{If it defines a relation variable matched by the current stream 
			element}
   				\If{$\exists \ (v,v')\in E_p: M_p(v)=m\land M_p(v') = m' 
   				$}{\label{line:subgraphif}
   					$\Lambda \gets \Lambda \cup \mathit{BFSmatch}(G^{(t)},I^{(t)}_{in}, 
   					I^{(t)}_{out},p,{s})$\label{line:getMatch}\;
   				}
   			}
   			$t' \leftarrow t$\;
      }
    \Return{$\Lambda$}\;
  \end{algorithm}
\end{minipage}
\vspace{-11pt}
\end{figure} 

\sstitle{Complexity}
For each stream element ${s} \in \mathcal{S}$, the most 
time-consuming operation of the above matching algorithm is the function $BFSmatch()$. It 
takes $\mathcal{O}(|p|_{max} \cdot d_{max})$ time, 
where 
\begin{equation}
|p|_{max} = \max_{(V_p,E_p, M_p) \in P} |V_p|
\end{equation}
is the size of the largest pattern in $P$, and 
\begin{equation}
d_{max}= \max_{v\in V^{(t)}} |\{ v'\in V^{(t)}\mid (v,v')\in E^{(t)} \}|
\end{equation}
is the maximum degree over all entities in the social graph. Processing 
${s}$ thus takes $\mathcal{O}(|P| \cdot 
|p|_{max} \cdot d_{max})$ time in total. Since these parameters are 
comparatively small, the runtime 
of \autoref{alg:match} is asymptotically constant and can be estimated 
empirically.

Note that a stream element $s$ is unique (i.e., there cannot be more than one 
relation between two entities) and timestamped at $t$.  Therefore, at time $t$, 
it can generate at most one new match for one specific pattern.

\subsection{Streaming Anomaly Computation}
\label{sec:incremental_anomaly}

Recall that anomaly-based rumour detection as discussed in 
\autoref{sec:anomaly_detection} relies on the matches of rumour patterns and a 
predefined set of features. Based thereon, local 
anomalies related to individual entities and relations are derived, before 
they are aggregated to global anomalies that span a whole subgraph. 
Having 
discussed how pattern matching is conducted in an online manner, we now present 
a streaming 
approach for the computation 
of local, element-level anomaly scores as well as of global, subgraph-level 
anomaly scores.

\sstitle{Streaming element-level scoring}
In a streaming setting, it is infeasible to maintain all historic data. 
However, even if it would be possible, the re-computation of anomaly scores 
every time new data arrives would not scale to the volume of the data emitted 
by social networks. We therefore propose a bounded approximation of local 
anomaly scores that works online.

Recall that the computation of local anomalies is based on a predefined set of 
features, as summarised in \autoref{sec:features}. Comparing the
current and past values of a feature vector assigned to a single entity or 
relation then enables the computation of a local anomaly score, see 
\autoref{sec:local_global}. In the absence of the historic values of a feature, 
one may assume that the historical values follow a particular distribution, 
e.g., a Gaussian one. Then, the mean and standard deviation of this 
distribution could be maintained in constant time and memory. At a specific 
point in time, a feature value could be declared to be an anomaly, if 
it is higher than a specified threshold, e.g., the 95\% percentile. 

However, the above approach has an important downside. It requires an 
assumption on the underlying distribution and, hence, may lead to incorrect 
anomaly scores if that distribution is not known. We therefore resort to an 
approach that imposes solely a weaker assumption. That is, we assume that the 
mean, the expected value of the feature, at the current point in time is the 
same as the earlier mean, i.e., before the current point in time. 

Conceptually, we divide the historic values of a feature into two classes: 
those of the current time ($t$) and those of earlier points in time ($<$$t$). 
Practically, we maintain two Count-Min-Sketch (CMS)~\cite{cormode2005improved} 
data structures to capture the evolution of the values of a feature $f$ for 
an element (entity or relation) $x$ of the social graph. The 
first CMS shall approximate the sum of the feature values up to the 
current time, denoted by $s^{(t)}$. 
At any time, the approximation of this 
sum, denoted by $\hat{s}^{(t)}$, can be obtained in constant time and 
memory. The second CMS is used to capture the feature value at the current time, 
denoted by $f^{(t)}$, while the approximated value is 
$\hat{f}^{(t)}$. 
Here, the existence of the second CMS is motivated by the 
fact that many features lend themselves for online computation. 
For example, to compute the number of retweets of the tweet represented by 
entity $v_3$ in 
\autoref{fig:rumour}, all paths of from $v_3$ to a user entity and, 
subsequently, to a tweet entity must be identified. 
Instead of counting all these paths repeatedly, at any time instant, it is more 
efficient to maintain the CMS and rely on the approximated value. 

As a result, the 
local anomaly score, i.e., the empirical p-value, that quantifies the number of 
past values that differ from the current observation, is computed as the 
chi-square statistic, capturing the difference between the observed ($obs$) and 
expected values ($exp$), of the two classes:

\small
\begin{equation}
\label{eq:p-value_feature}
\begin{split}
\hat{p}(f,x)^{(t)} &= X^2\\ 
&= \frac{(obs^{(t)} - exp^{(t)})^2}{exp^{(t)}} + \frac{(obs^{(<t)} - 
exp^{(<t)})^2}{exp^{(<t)}}\\
&= \frac{(\hat{f}^{(t)} - \hat{s}^{(t)}/t)^2}{\hat{s}^{(t)}/t}  + \frac{( 
\hat{s}^{(t)} - \hat{f}^{(t)} - \hat{s}^{(t)}(t-1)/t)^2}{\hat{s}^{(t)}(t-1)/t}\\
&= \left(\hat{f}^{(t)} - \frac{\hat{s}^{(t)}}{t}\right)^2 \ 
  \frac{t^2}{\hat{s}^{(t)}(t-1)}
\end{split}
\end{equation}

\normalsize
\noindent
The above formulation enables us to derive a bound for the approximation of the 
anomaly score, as follows.

\begin{myprop}[Confidence of streaming p-value]
\label{prop:bound}
  If a feature value with $\hat{p} < \alpha$, with $0 < \alpha < 1$, is 
  declared to be an anomaly, the confidence level of this conclusion is 
  $1-\alpha$.
\end{myprop}

While the above p-value is computed for each feature of an element (entity or 
relation) of a social graph, the anomaly score of the element is derived by:
\begin{equation}
\label{eq:sum_p}
	p(x)^{(t)} = \frac{1}{t-1} \sum_{t'=1}^{t-1} \mathds{1}_{p_{min}(x)^{(t')} \leq p_{min}(x)^{(t)}}
\end{equation}
where $\mathds{1}_{(.)}$ equals to one if the condition $(.)$ is satisfied, and 
zero otherwise. And $p_{min}(x)^{(t')} = \min_f p(f,x)^{(t)}$ is the minimum 
p-value across all features of $x$. 
Following~\cite{tam2019anomaly}, $min$ is used as an aggregation function to  
avoid false positives, where some features indicate an anomaly, whereas other 
do not. Moreover, we do not consider the minimum p-value over all features at a 
single timestamp directly, since elements can have different numbers of 
features. Rather, our idea is to cross-check the scores between different 
timestamps across features, so that our aggregation yields uniform scores over 
all entities and relations, regardless of their modality.

Again, we adopt CMS data structures and chi-square statistics to compute 
$p(x)^{(t)}$ efficiently. Instead of maintaining the value $p_{min}(x)$ at all 
time points, we consider the p-value for each feature of $x$ 
(\autoref{eq:p-value_feature}) as streaming input. Using a similar mechanism as 
above, we obtain the approximate p-value, denoted by $\hat{p}(x)^{(t)}$. To 
derive a bound for this approximation, we can use the union bound (Boole's 
inequality) on \autoref{eq:sum_p}, resulting in a $1 - \alpha / (t - 1)$ 
confidence. In other words, the more historical data we have, the more 
confidence one may put in our anomaly scores.



\sstitle{Streaming subgraph scoring}
Based on the local anomaly scores per element (entities and relations) of a 
social graph, global anomaly scores are computed for connected subgraphs. To 
this end, it was suggested to rely on scan 
statistics~\cite{kulldorff1997spatial} that incorporates all elements with 
anomaly scores that are significant at a specific confidence level. 
As detailed in~\cite{tam2019anomaly}, such an approach leads directly to an 
online procedure to approximate the respective subgraphs. In essence, it 
leverages the fact that newly added nodes may directly be characterised as 
being rumour related, either due to being connected to an existing rumour in 
the social graph or due to inducing a new rumour subgraph. 

\subsection{Complete View on Streaming Rumour Detection}
\label{sec:incremental_integration}

Finally, we discuss how pattern matching and anomaly scoring as introduced 
above are combined for streaming rumour detection. Considering 
\autoref{alg:match}, the anomaly scoring is incorporated as follows. First, the 
element-level scoring is conducted immediately once a new stream element ${s} = 
(u,u',m,m',t) \in \mathcal{S}$ is received (\autoref{line:for}). That is, the 
p-value is computed for each feature of entity $u$, entity $u'$, and relation 
$(u,u')$ (using \autoref{eq:p-value_feature}). Second, the matches of query 
patterns are only collected if their anomaly scores have a confidence value 
above of specific threshold (typically 95\%). That is, for each pattern match 
obtained by $\mathit{BFSmatch}(G^{(t)},I^{(t)}_{in}, 
I^{(t)}_{out},p,{s})$ (\autoref{line:getMatch}) the global anomaly score is 
computed and used to decide whether the respective subgraph shall be added to 
the result set $\Lambda$. This way, only the pattern matches that denote actual 
rumours are returned by the algorithm. 

%
%
%
%
%
%
%
\section{Load Shedding}
\label{sec:shedding}

Next, we turn to the load shedding mechanism in our framework for 
efficient stream rumour detection, see \autoref{fig:framework}. 
\edit{However, doing so is not straight-forward as rumour detection is now streaming. Seemingly unimportant data elements might turn out to be important later and vice-versa.}
We first 
formulate the goal of load shedding as an optimization problem, i.e., we aim at 
guaranteeing a specific latency of processing while minimising the loss in the 
detection coefficient (\autoref{sec:shedding_problem}). To address 
this problem, we introduce a coefficient model to capture the importance 
of data elements in the input buffer (\autoref{sec:coefficient}). Then, we 
present strategies to decide when, how much, and what to shed based on this 
model (\autoref{sec:algorithm}). Finally, we discuss extensions of this general 
approach that cope with more complex application scenarios 
(\autoref{sec:extension}).

\subsection{Problem Formulation}
\label{sec:shedding_problem}

When a streaming system becomes overloaded, i.e., the input rate is larger than 
the processing rate, load shedding is needed to keep the processing latency 
low~\cite{HeBN14}. Shedding of elements of a data stream is particularly useful 
in rumour detection: Dropping data about entities and relations that have a low 
probability of being part of a match of a rumour pattern paves the way for the 
detection of actual rumours, which can be mitigated when detected early. 

However, load shedding potentially leads to a loss of matches of rumour 
patterns and, hence, may lower the detection coefficient. Depending on the 
actual approach employed to detect rumours based on pattern matches, the 
consequence of load shedding may be both, 
false negatives, i.e., missed rumours, as well as false positives, i.e., 
falsely detected rumours. We illustrate this aspect with the following example. 

\begin{example}
	Two patterns from \autoref{fig:pattern} yield matches for the social graph 
	constructed for \autoref{fig:rumour} at time point $5$. That is, for 
	pattern $p_1$, we observe a match in which the entity variables of $p_1$ 
	are mapped to the entities $v_3$, $v_4$, $v_5$, and $v_6$. Moreover, for 
	pattern $p_2$, there is a match that maps the entity variables to entities 
	$v_1$, $v_3$, $v_6$, and $v_7$, as discussed already in 
	\autoref{ex:pattern_match}. Dropping information about the relation 
	modelled by edge $e_5 = (v_3,v_6)$ would eliminate these two matches, which 
	may lead to false negatives: Rumours identified based on these patterns 
	could be missed. On the other hand, the loss of edge $e_6 = (v_6, v_7)$ 
	might lead to false positives. That is, if the user modelled by the entity 
	$v_6$ creates another tweet at a later point in time (\textgreater $5$), a 
	match of 
	$p_2$ would be induced. As such, a rumour may falsely be identified based 
	on the later tweet.
\end{example}

Effective load shedding shall drop the elements of the data stream with the 
least impact on the detection coefficient. As such, our objective is to 
minimise the relative difference in the detection coefficient between rumour 
detection with and without load shedding, while satisfying a latency bound in 
the case of shedding.

\begin{myprob}[Load Shedding in Streaming Rumour Detection]
  \label{problem}
  Given a stream prefix $\mathcal{S}(t)$, a set of patterns~$P$, and a rumour 
  detector $\Gamma$, 
  the problem of \emph{load shedding in streaming rumour detection} is to 
  design a shedding function $\rho$, such that the latency of processing any 
  stream element $s \in \mathcal{S}(t)$ stays below a threshold $\theta$ and 
  the following 
  coefficient loss is minimal:
  \begin{equation}
  \label{eq:coefficient_loss}
    \frac{f(\Gamma, P, \mathcal{S}(t)) - f(\Gamma, P, \rho(\mathcal{S}(t))) 
    }{f(\Gamma, P, \mathcal{S}(t))}	.
  \end{equation}
  where $f(\Gamma,P,\mathcal{S}(t))$ is the detection 
coefficient, which is defined as the number of rumours detected by the rumour 
detector $\Gamma$ over stream prefix $\mathcal{S}(t)$ based on the set of 
rumour patterns $P$.
\end{myprob}

In general, \autoref{eq:coefficient_loss} represents the degradation of rumour detection performance due to load shedding to preserve the processing latency.
The above problem becomes more challenging when $t$ is not fixed to a single 
time point, but considered over a large time interval. Since 
real-world rumours are dynamic and have different characteristics over time, 
rumour detection faces concept drift. As a consequence, a shedding strategy 
that was effective in the past may become obsolete in the future.

\subsection{Coefficient Modelling}
\label{sec:coefficient}

To avoid shedding elements of a data stream that contribute significantly to 
matches 
of rumour patterns, 
we assess the contribution of stream elements using a coefficient model. The 
model is based on the modalities defined by the stream element for a pair of 
entities (and, hence, a relation), as well as their relative position within 
the windows in which the data is emitted by a social network, see 
\autoref{sec:model}.
These choices are motivated 
by the following considerations: (i) the model shall be lightweight, i.e., 
it should comprise only relatively primitive information of the stream; (ii) 
the modalities determine to which extent entities and, hence, relations 
participate 
in matches of rumour patterns; (iii) the closer entities and, hence, relations, 
that are part of pattern matches in a stream, the more likely the matches 
denote actual rumours, since it is well-known that rumours propagate 
fast~\cite{vosoughi2018spread}.


\sstitle{Statistical coefficient}
Based on the above considerations, we build a statistical model for the 
detection coefficient. That is, we count the number 
of occurrences of each second-order modality of the relation described by a 
stream element (which implicitly covers the entities) that contribute to the 
detected rumours per position in a 
window. 
We tune the granularity of this model by normalizing these counts to a 
predefined interval, given as $[0,100]$ in the remainder.
The coefficient values are stored in a coefficient matrix 
$\Pi_{|\mathcal{M}^2| \times |w|}$, where $|\mathcal{M}^2|$ is the number of 
second-order modalities and $|w|$ is the window size. A cell 
$\Pi((m,m'),i)$ of the matrix captures the coefficient value, normalized to 
$[0,100]$, for a stream 
element $s= (u,u',m,m',t) \in \mathcal{S}$ at the position $i$ of a window 
defined over the stream.

\sstitle{Shedding by coefficient thresholding}
Load shedding shall drop the $k$ relations with the lowest coefficient 
values in a window (we later discuss how to choose a value of $k$). 
Yet, a naive realisation of such a shedding function would have several 
drawbacks. It would require to wait until the window is fully available 
(which implies a $\mathcal{O}(|w|)$ wait time), to then sort the relations 
(in $\mathcal{O}(|w| 
\log k)$, e.g., by heap sort), while the overhead would be induced for each 
window.

Against this background, we consider a more efficient approach that relies on a 
coefficient threshold, denoted by $\pi_{min}$, to drop the desired number of 
relations on-the-fly. Here, the challenge is, given a value for~$k$, to set 
$\pi_{min}$ accordingly, so that exactly $k$ relations are actually dropped.
We address this challenge based on the properties of cumulative 
distribution functions. That is, for a given window~$w$, we define the number 
of occurrences of a second-order modality, 
for which the coefficient is less or equal to a certain value $\pi$ as follows:

\small
\begin{equation}
\Omega(\pi) = |\{(m,m') \in \mathcal{M}^2 \mid 
\sum_{1\leq i \leq |w|} \Pi((m,m'),i) \leq \pi\}|.
\end{equation}

\noindent
\normalsize
We refer to $\Omega$ as the cumulative coefficient occurrence 
(CCO). Based thereon, the coefficient threshold $\pi_{min}$ is derived using 
the inverse function of CCO for a given the number of relations $k$, i.e.,  
$\pi_{min} = \Omega^{-1}(k)$.


The CCO $\Omega$ can be derived from the coefficient matrix $\Pi$ using dynamic 
programming, as shown in \autoref{alg:utility}. Here, based on each window 
position and each second-order modality, we first compute $\omega_\pi$, the 
number of occurrences of each individual coefficient value $\pi \in [0,100]$. 
Here, the division by $|\mathcal{M}^2|$ (\autoref{l:m2}) is explained by the 
fact that a single window position $i$ leads to incremented occurrences of 
coefficients of multiple second-order modalities; 
and hence, $\omega_\pi$ is shared by $|\mathcal{M}^2|$ second-order modalities.
Subsequently, the CCO is constructed by accumulating the respective numbers of 
occurrences of individual coefficient values $\omega_\pi$.
\autoref{alg:utility} has a runtime complexity of 
$\mathcal{O}(|\mathcal{M}^2| 
\times |w| + 
100)$, which can be considered as a constant 
update time.

\begin{figure}[t]
  \begin{minipage}[t]{1.0\linewidth}
    \removelatexerror
    \begin{algorithm}[H]
      \footnotesize
      \caption{Coefficient modelling}
      \label{alg:utility}
      \SetKwInOut{Input}{input}
      \SetKwInOut{Output}{output}

      \Input{
        Current occurrence statistics $\Pi$
      }
      \Output{
      Cumulative coefficient occurrence $\Omega$
      }
	\BlankLine      
      \For{$i\gets 1..|w|$}{
      		\For{$ (m,m') \in \mathcal{M}^2$}{
      			$\pi \gets  \Pi((m,m'),i)$\;      			
      			$\omega(\pi) \gets \omega(\pi) + 
      			1/|\mathcal{M}^2|$\label{l:m2}\;
      		}
      }

	$\Omega(0) \gets \omega(0)$\;
	
	\lFor{$\pi\gets 1..100$}{
		$\Omega(\pi) \gets \omega(\pi) + \Omega(\pi-1)$}
      
    \Return{$\Omega$}
  \end{algorithm}
\end{minipage}
\vspace{-10pt}
\end{figure}

\sstitle{Handling concept drift}
The distribution of streaming data may show concept drift, i.e., it may be  
non-stationary~\cite{yu2018request,knoblauch2018doubly}. In social networks, 
the propagation structure of rumours may change over time, which affects 
the coefficient distribution of second-order modalities, i.e., the coefficient 
matrix. As such, the 
model may become outdated and needs to be retrained. 

To determine the need for 
retraining, we monitor changes to the coefficient matrix and recompute the CCO 
once a drift is detected. Specifically, we rely on the mean relative error 
(MRE)~\cite{peierls1935statistical} to compare the current coefficient matrix 
with the one used to derive the CCO. Using the MRE has the advantage 
that the measure can be updated online upon the arrival of a new stream 
element in $\mathcal{O}(1)$. 
Once the current coefficient matrix deviates by certain percentage from the one 
used to derive the current CCO, the model is recomputed using 
\autoref{alg:utility}.


\subsection{Load Shedding}
\label{sec:algorithm}

We are now ready to answer three fundamental 
questions for the design of a load shedding mechanism:

\sstitle{When to shed}
Upon the arrival of a stream element, the matching of rumour patterns needs to 
consider all stream elements that are currently in the input buffer, which are 
potentially part of multiple 
windows. As such, the latency of 
processing an arriving stream element is given as $t_{match} \times b$, 
where $b$ is the number of stream elements currently in the buffer and 
$t_{match}$ is 
the delay induced per element. According 
to \autoref{problem}, we must ensure that $t_{match} \times b \leq \theta$.
Hence, we monitor the input queue size and perform load shedding when:
\begin{equation}
\label{eq:condition}
b > \alpha \times \frac{\theta}{t_{match}}
\end{equation}
where 
$\alpha \in [0,1]$ is a trade-off factor. 
If we perform load shedding too early (i.e., a small value for $\alpha$), new 
stream elements are more likely to be dropped, especially in case of bursts 
that increase the latency for a short time period. 
If we perform load shedding relatively late (a large value for $\alpha$), 
elements with high utility are more likely to be shed as the coefficient 
threshold would be higher. Therefore, we update $\alpha$ dynamically according 
to the current distribution of coefficients. The value of $\alpha$ is selected 
such that a new window $w$ is likely to have at least $k$ 
elements with low utility, i.e., $\alpha = {k}/{|w|}$.

\sstitle{How much to shed}
A system is overloaded if the input rate $r$ (which is dynamic) is strictly 
larger than the processing rate $r_{match}$ (which is static). Moreover, a 
rumour detector $\Gamma$ can be assumed to have a constant update 
time~\cite{zubiaga2018detection}. In an overload situation, there will be an 
extra $r-r_{match}$ number of stream elements in the input buffer per second, 
which increases the latency of processing a new element.
Therefore, we must drop $k$ elements in the current window to compensate for 
the additional elements. With $b$ as the number of elements currently in the 
buffer and $b_{max}$ as its size, we derive the number of elements to 
shed, $k$, as follows:
\begin{equation}
\label{eq:amount}
k = \max\left(\frac{r-r_{match}}{r} \times |w|, b - b_{max} + |w|\right).
\end{equation}
The second term of the $\max$ function is explained by the fact that when 
pushing the remaining stream elements of a window to the input buffer, its size 
must not be exceeded. That is, we need to ensure that the condition 
$b+|w|-k\leq b_{max}$ is satisfied.

\sstitle{What to shed}
As mentioned, we drop stream elements with coefficients smaller than 
or equal to a coefficient threshold $\pi_{min}$.
To set $\pi_{min}$, we iterate over an array representation of the cumulative 
coefficient occurrence $\Omega$ 
(which is sorted by design) and define $\pi_{min}$ based on the index of the 
first element with $\Omega[\pi] \geq k$:
\begin{equation}
\label{eq:threshold}
\pi_{min} = \argmin_{\pi \in [0,100]} \{\Omega(\pi)\mid \Omega(\pi) \geq k \}
\end{equation}

\begin{figure}[t]
	\begin{minipage}[t]{1.0\linewidth}
		\removelatexerror
		\begin{algorithm}[H]
			\footnotesize
			\caption{Load shedding procedure}
			\label{alg:shed}
			\SetKwInOut{Input}{input}
			\SetKwInOut{Output}{output}
			
			\Input{
				A data stream $\mathcal{S}$ with windows of size $|w|$;\\
				\ the size of the input buffer $b_{max}$.
			}

			\BlankLine
			$w_{\mathit{current}}\gets \emptyset$\tcp*{Current 
			window}\label{alg:init_1}
			$\mathit{shedding} \gets \bot$\tcp*{Shedding flag}
			$\pi_{min}\gets 0$\tcp*{Coefficient threshold}
			$\mathcal{S}' \gets \emptyset$\tcp*{Stream after shedding}
			\label{alg:init_2}

			\BlankLine
			\For{${s} = (u,u',m,m',t) \in \mathcal{S}$ ordered by $t$}{
				\label{alg:per_element}
				$k, \alpha\gets 0$\label{alg:shed_par}\tcp*{Shedding 
				parameters} 
				\If{time-based measurement interval elapsed}{
					\label{alg:measurement_1}
					\tcp{Set shedding parameters dynamically:}
				
					Set $k$ based on $r$, $r_{match}$, $|w|$, $b$ using 
					\autoref{eq:amount}\;
					$\alpha \gets {k}/{|w|}$\label{alg:measurement_2}\;
				}
			
				\BlankLine
				\tcp{Is load shedding active?}
				\If{$\mathit{shedding} = \top$}{
					\label{alg:shedding_active}
					$i\gets 
					\left|\{(\bar{u},\bar{u}',\bar{m},\bar{m}',\bar{t})\in 
					w_{\mathit{current}} \mid \bar{t}\leq t  
					\}\right|$\label{alg:shedding_1}\;
					\tcp{Keep elements not shed (realise $\rho$)}
					\lIf{$\Pi((m,m'),i) > \pi_{min}$}{
						$\mathcal{S}' \gets \mathcal{S}'\cup \{s \}$
						\label{alg:shedding_2}
					}
				}
				\BlankLine
				$w_{\mathit{current}}\gets w_{\mathit{current}} \cup \{ 
				s\}$\label{alg:current_window}\tcp*{Extend window}
				\If(\tcp*[f]{Window complete?}){$|w_{\mathit{current}}| = 
				|w|$\label{alg:last_element}}{
					$w_{\mathit{current}}\gets 
					\emptyset$\label{alg:reset_window}\tcp*{Reset current 
					window}
					\tcp{Is shedding needed according to 
					\autoref{eq:condition}?}
					\If{$b > \alpha \times {\theta}/{t_{match}}$}{
						\label{alg:needed}
						\tcp{Coefficient threshold for filtering}
						Set $\pi_{min}$ using 
						\autoref{eq:threshold}\label{alg:threshold}\;
						$\mathit{shedding} \gets 
						\top$\label{alg:act}\tcp*{Activate shedding}
					}
					\lElse(\tcp*[f]{Deactivate 
						shedding}){$\mathit{shedding} \gets 
						\bot$\label{alg:deact}}	
				}
				\BlankLine
				Pattern matching and rumour detection using 
				$\mathcal{S}'$\label{alg:pattern_matching}\;					
			}
		      
		\end{algorithm}
	\end{minipage}
	\vspace{-10pt}
\end{figure}

\noindent
\autoref{alg:shed} formalises the load shedding procedure, which encapsulates 
the actual pattern matching and rumour detection. It takes as input a data 
stream $\mathcal{S}$ over which (non-overlapping) windows of size $|w|$ have 
been defined, along with information on the size of the input 
buffer, $b_{max}$. The algorithm first initialises several auxiliary variables 
(lines~\ref{alg:init_1}-\ref{alg:init_2}), i.e., the current window 
($w_{current}$), a flag that indicates whether load shedding has been activated 
($\mathit{shedding} \in \{\top,\bot \}$), the coefficient threshold 
($\pi_{min}$), and the result stream without the elements that have been shed 
($\mathcal{S}'$).

It then processes each element of the stream (\autoref{alg:per_element}). The 
initial shedding parameters, i.e., $k$, the number of elements to shed and 
$\alpha$, the trade-off factor, capture the absence of shedding 
(\autoref{alg:shed_par}). Upon some time-based 
trigger, these parameters are updated dynamically
(lines~\ref{alg:measurement_1}-\ref{alg:measurement_2}). That is, $k$ is 
determined based on the input rate, the processing rate, and the current number 
of elements in the input buffer (see \autoref{eq:amount}), while $\alpha$ is 
derived from $k$ and the window size $|w|$.

If load shedding is active (\autoref{alg:shedding_active}), the algorithm 
checks based on the coefficient matrix whether the current element shall either 
be shed or kept and, hence, added to the result stream 
(lines~\ref{alg:shedding_1}-\ref{alg:shedding_2}). We thereby implicitly 
realise the load shedding function $\rho$, as defined in \autoref{problem}.

In any case, the current element is recorded as part of the current window 
(\autoref{alg:current_window}). If it was the last element of the window 
(\autoref{alg:last_element}), the window is reset 
(\autoref{alg:reset_window}) and the status of the 
system is assessed (\autoref{alg:needed}). If the system is overloaded, the 
coefficient threshold $\pi_{min}$ is computed (\autoref{alg:threshold}) and 
load shedding is activated (\autoref{alg:act}). Otherwise, load shedding is 
deactivated (\autoref{alg:deact}). 
Finally, pattern matching and rumour detection are conducted based on the 
result stream $\mathcal{S}'$, from which some elements may have been shed 
(\autoref{alg:pattern_matching}). 

\subsection{Extensions}
\label{sec:extension}

Having introduced our general approach to load shedding in stream rumour 
detection, we discuss how to cope with more complex application scenarios.

\sstitle{Variable window sizes}
We lift our approach to windows of variable sizes, 
see~\cite{slo2019espice}, through two adaptations:

\begin{compactitem}
	\item \emph{Coefficient Modelling:} If windows have variable size, $|w|$ no 
	longer serves as the number of 
	columns for the coefficient matrix. However, in practice, differences in 
	window sizes can be expected to be relatively modest. Therefore, we 
	consider an estimated window size 
	$\overline{|w|}$, which, for instance, corresponds to the maximal size of 
	windows seen so far. Then, the relative 
	position of a stream element in a window $w_i$ is scaled by a factor of 
	${|w_i|}/{\overline{|w|}}$ to the coefficient matrix $\Pi_{|\mathcal{M}^2| 
	\times \overline{|w|}}$.
	\item \emph{Load Shedding:} With variable window sizes, the size of the 
	current window may be unknown until its end. Consequently, we cannot use 
	\autoref{eq:amount} to compute how many stream elements shall be dropped. 
	Again, the solution is to rely on an estimated window size 
	$\overline{|w|}$. However, more precise 
	methods to derive the number of elements to shed based on window size 
	prediction can be found in~\cite{slo2019espice}.
\end{compactitem}

\sstitle{Variable shedding intervals}
Once shedding is triggered according to \autoref{eq:condition}, it may be 
reasonable not to drop stream elements immediately. The reason being that the 
coefficient values may not be evenly distributed, e.g., elements with high 
utility may all be located in a certain range of the stream. Hence, dropping 
many stream elements consecutively potentially amplifies the coefficient loss, 
although it reduces the processing latency.

Against this background, we observe that the input buffer, in general, has 
spare space of size $z = \frac{\theta}{t_{match}} - \alpha \ 
\frac{\theta}{t_{match}}$. If the current 
window is smaller than that, $|w_{current}| \leq z$, one may wait for the next 
window to fill up before taking any shedding decision.
However, if $|w_{current}| > z$, there is a risk of a latency violation if 
high-utility stream elements arrive consecutively, which means that more than 
$z$ elements can be pushed to the input buffer 
before the window ends. 
To handle this situation, we divide the current window into smaller 
parts of size $v = \lfloor z \rfloor$ and perform load shedding for these 
parts, using $|v|$ instead of the window size $|w|$ in \autoref{eq:amount}.

\section{Empirical Evaluation}
\label{sec:exp}

We evaluated our approach with a dataset of rumours on Twitter. We first 
discuss the experimental setup (\autoref{sec:exp_setup}), before turning to an 
evaluation of the following aspects of our approach:
\begin{compactitem}
	\item The effectiveness of our shedding strategies (\autoref{sec:exp_accuracy_shedding}).
	\item The effectiveness of our approach to streaming rumour detection 
	(\autoref{sec:exp_accuracy_detect}).
	\item The efficiency of our shedding strategies (\autoref{sec:exp_time_shedding}).
	\item The efficiency of streaming rumour detection 
	(\autoref{sec:exp_time_detect}).
	\item The robustness of the presented techniques 
	(\autoref{sec:exp_sensitivity}).
	\item The end-to-end performance of our framework based on ablation tests 
	(\autoref{sec:exp_ablation}).
\end{compactitem}

\subsection{Experimental Setup}
\label{sec:exp_setup}

\sstitle{Datasets}
We collected data comprising 4 million tweets, 3 million users, 28893 
hashtags,
and 305115 linked articles, including around 1022 rumours~\cite{tam2019anomaly}.
The data spans several domains, each of which is a separate dataset:

\begin{compactitem}
	\item \emph{Politics:} rumours related to political issues.
	\item \emph{Crime:} rumours related to incidents such as 
	the 2017 Las Vegas shooting.
	\item \emph{SciTech:} rumours related to scientific myths and exaggerated technological inventions.
\end{compactitem}

\sstitle{Rumour detection}
We consider various techniques for rumour detection:
\begin{compactitem}
	\item \emph{Ground:} The rumour detector is simulated by the ground-truth 
	itself.
	\item \emph{Decision}~\cite{castillo2011information}: A decision
	tree classifier that is based on the Twitter information credibility model.
	The decision tree is constructed based on several hand-crafted features.
	\item \emph{Nonlinear}~\cite{yang2012automatic}: An SVM-based approach that 
	uses a set
	of hand-crafted features, selected for the tweets to classify.
	\item \emph{Rank}~\cite{zhao2015enquiring}: A rank-based classifier that
	aims to identify rumours based on enquiry tweets.
	\item \emph{Static}~\cite{tam2019anomaly}: A static version of 
	anomaly-based rumour detection, which is based on 45 rumour patterns. 
	Instead of using the presented mechanisms for online anomaly computation 
	(\autoref{sec:incremental_anomaly}), it derives the anomaly scores from the 
	raw historical data. d
	\item \emph{Dynamic:} The streaming version of anomaly-based rumour 
	detection, as presented in this paper.
\end{compactitem}
Note that further, similar rumour detection techniques have been proposed in 
the literature, see~\cite{zubiaga2018detection}. Most of them are not 
applicable in our context, which prevents a fair comparison. The reasons for 
that are (i) that they require large domain-specific training data~\cite{ma2017detect,bian2020rumor}; 
(ii) that they employ domain-specific pre-processing steps (e.g., they segment 
the data according to some domain-specific 
heuristics~\cite{zubiaga2018detection}); and (iii) that they are often intractable in 
a streaming setting~\cite{nguyen2021judo}.

\sstitle{Metrics}
We complement common metrics for data streams~\cite{HeBN14}, such as 
\emph{throughput}, \emph{latency}, \emph{shedding ratio}  with the following 
evaluation measures:

\begin{compactitem}
	\item \emph{Shedding coefficient:} the ratio of detected rumours with 
	shedding and without shedding.
	\item \emph{$F_{\beta}$-score:} a weighted combination of true positives, 
	false negatives, and false positives~\cite{huang2017mention}. It measures 
	accuracy in the presence of imbalanced distributions. 
	In our case, the ratio of stream elements participating in a rumour and the 
	total number of elements is less than one. The weight $\beta$ is set to 
	this ratio for each dataset.
	\item \emph{Error rate:} the ratio of the stream elements that are part 
	of rumours and are shed, and their total number. 
\end{compactitem}

\sstitle{Shedding baselines}
We compare our approach with several baseline techniques for load shedding:
\begin{compactitem}
\item \emph{Random shedding:} this strategy sheds $k$ random elements of the 
current window, if the current input buffer satisfies $b > 
\frac{\theta}{t_{match}} + 1$ (the latter term ensures that processing of the 
last stream element in the buffer will have a latency $\leq \theta$).
\item \emph{Weighted shedding:} instead of uniform sampling $k$ stream elements 
to shed, this strategy weights the sampling probability of each element 
proportional to its coefficient.
\end{compactitem}

\sstitle{Environment}
All results have been obtained on an Intel i7 2.8GHz system (4 cores, 16GB 
RAM). We report 10-run averages to mitigate randomness. The default window size 
is $|w|=100$.

\subsection{Effectiveness of Shedding Strategies}
\label{sec:exp_accuracy_shedding}

We first evaluate the effectiveness of our shedding strategy, in the light of 
the aforementioned baseline techniques.

\sstitle{Accuracy}
We assess the accuracy of event detection, in terms of the 
$F_{\beta}$-score, for our shedding strategy in comparison to the baseline 
techniques, i.e., random shedding and weighted shedding. 


\begin{figure}[!h]
  \centering
  \vspace{-1.5em}
  \begin{subfigure}{.49\linewidth}
    \centering
    \includegraphics[width=1.\linewidth]{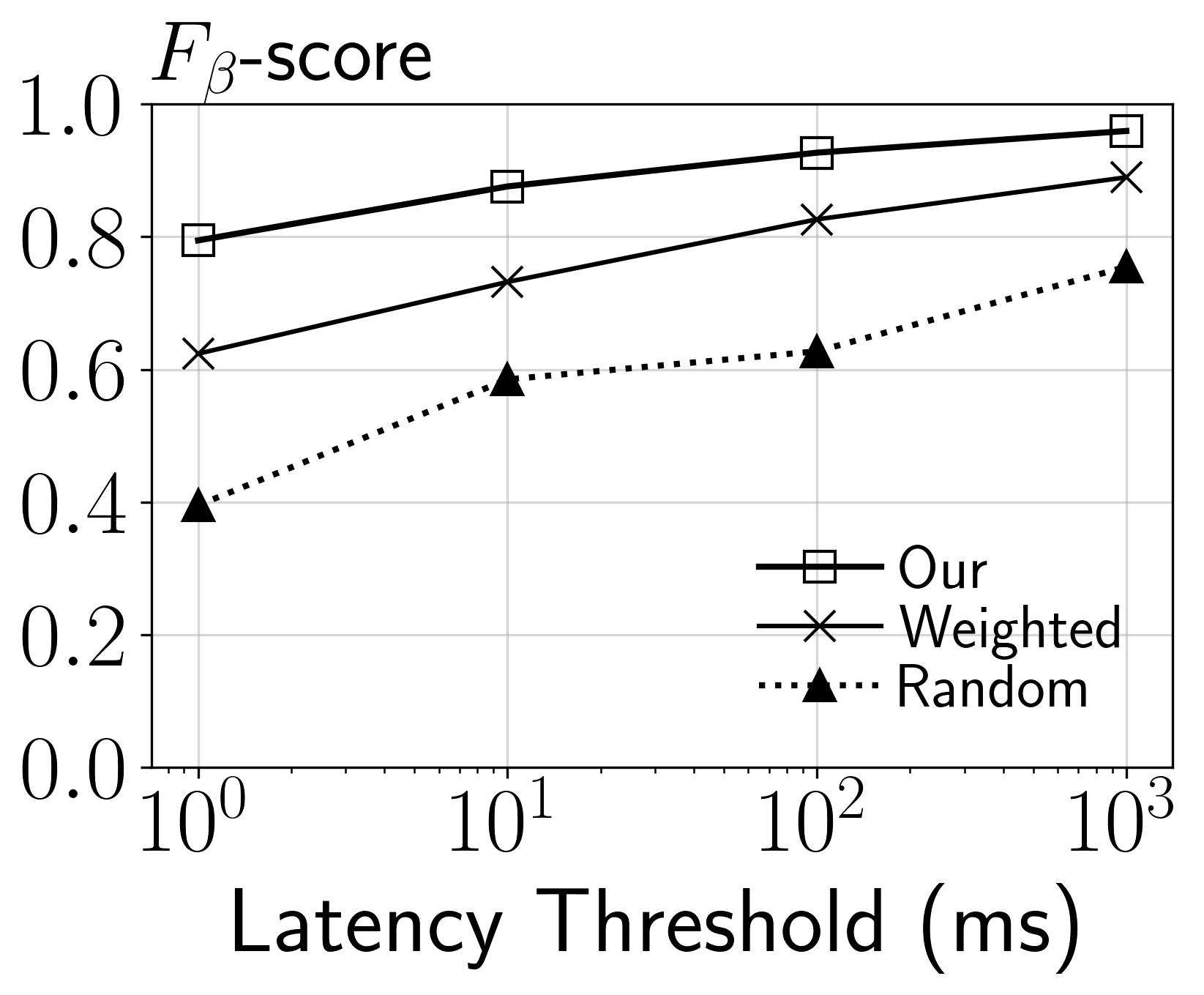}
    \caption{Politics dataset}
  \label{fig:exp_accuracy_politics}
  \end{subfigure}
  \begin{subfigure}{.49\linewidth}
    \centering
    \includegraphics[width=1\linewidth]{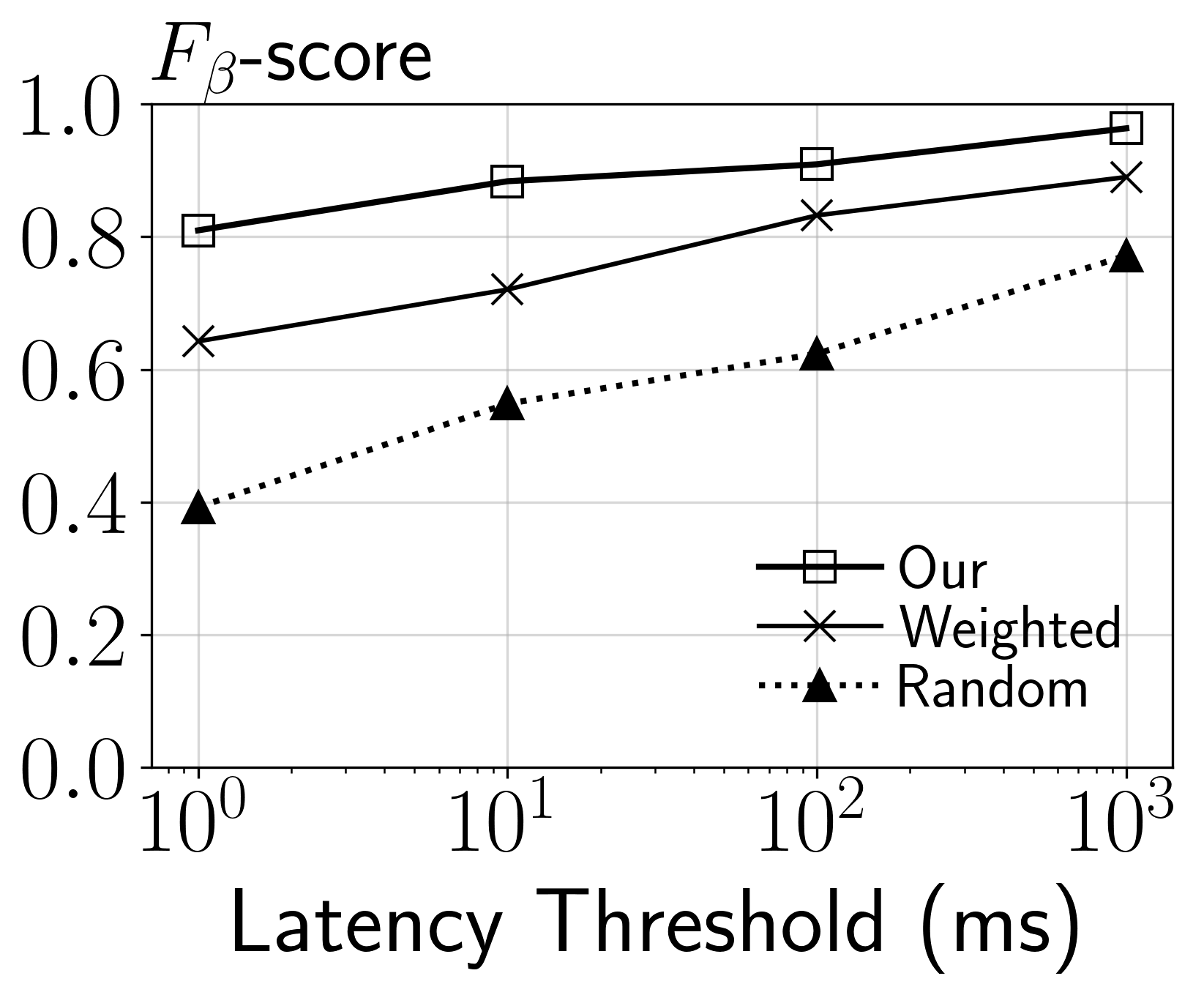}
    \caption{Crime dataset}
  \label{fig:exp_accuracy_crime}
  \end{subfigure}
  \begin{subfigure}{.49\linewidth}
    \centering
    \includegraphics[width=1\linewidth]{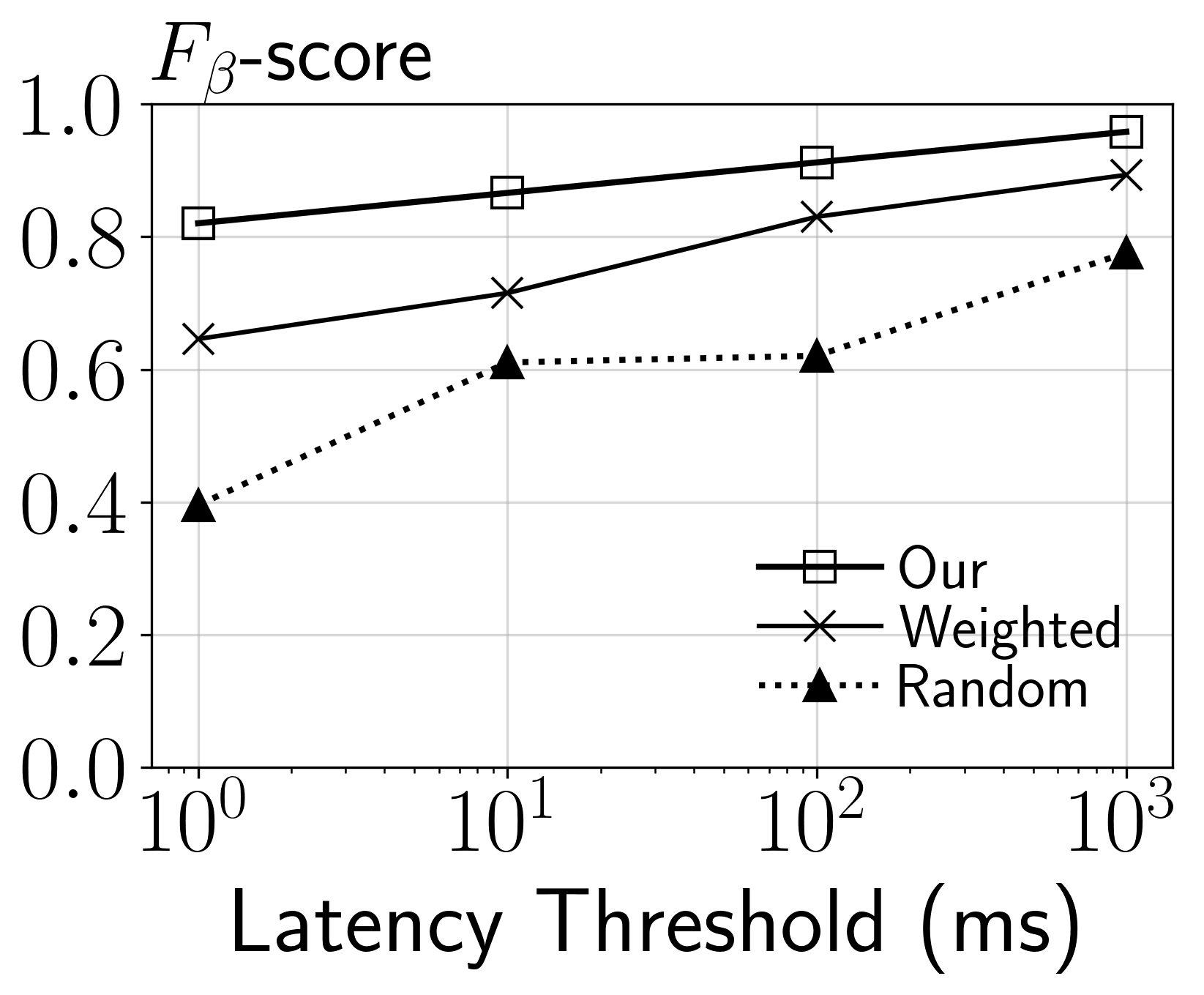}
    \caption{SciTech dataset}
  \label{fig:exp_accuracy_scitech}
  \end{subfigure}
  \caption{Accuracy of shedding strategies.}
  \label{fig:exp_accuracy}
\vspace{-1.5em}
\end{figure}

\autoref{fig:exp_accuracy} shows the results for each dataset. Here, 
the X-axis varies 
the latency threshold from $1s$ to $10^3s$ ($\approx 17min$), while the Y-axis 
reports the accuracy.
We observe that our method outperforms the baseline strategies. 
Also, when varying the latency 
threshold, better accuracy is obtained with larger bounds. This is expected 
since larger thresholds imply less shedding of stream elements.
Yet, the difference between the methods is also smaller with larger thresholds, 
as the shedding decisions are less critical once the overall amount of elements 
to shed is reduced. 

Note that the \emph{weighted} baseline performs 
worse than our method, even though both rely on the same coefficient 
information. The reason being that even elements with high coefficient values 
may be chosen by sampling. 

\begin{figure}[!h]
	\centering
	 \includegraphics[width=0.5\linewidth]{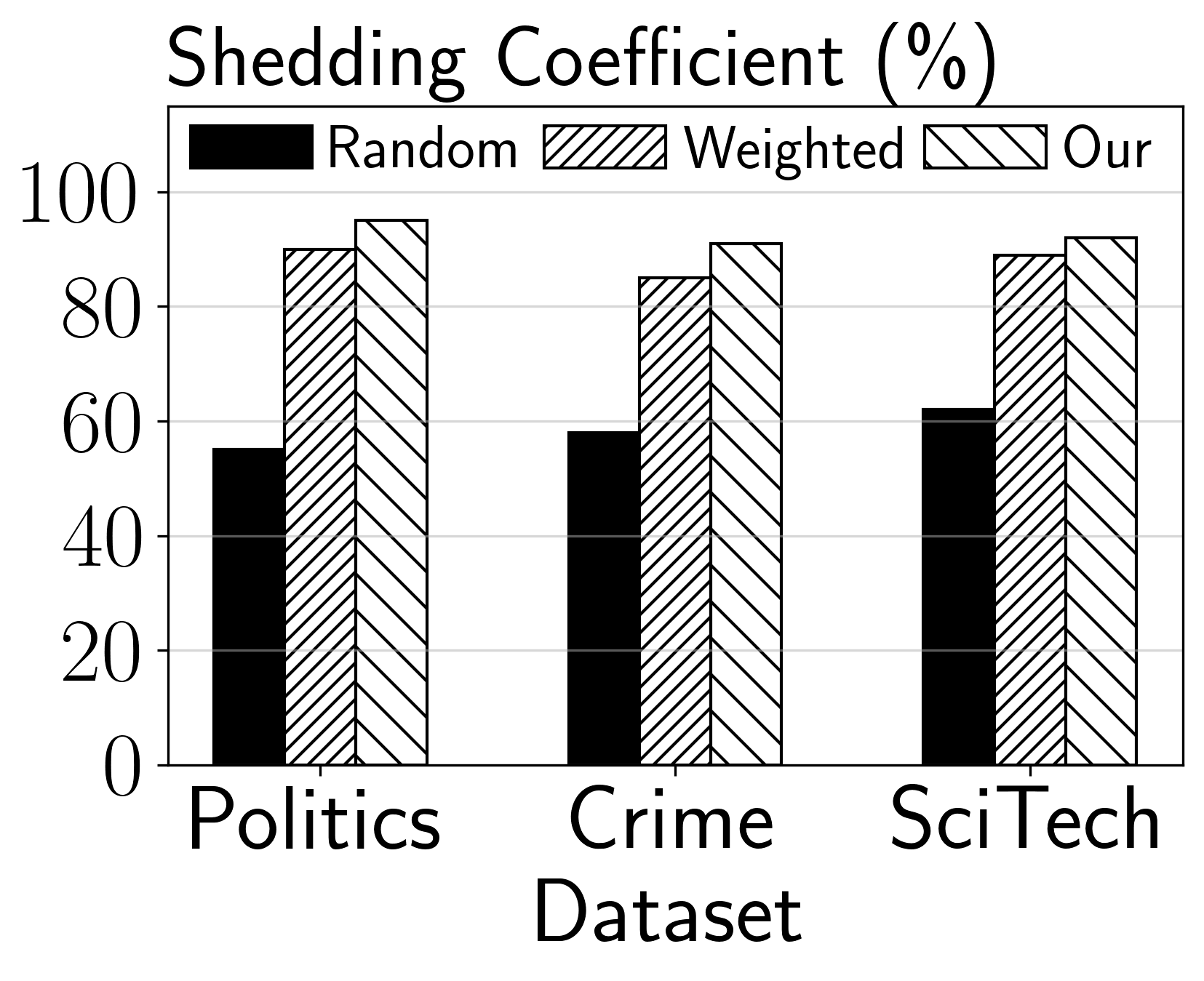}
	 \vspace{-5pt}
    \caption{Coefficient of shedding strategies.}
  \label{fig:exp_coefficient}
\end{figure}

\sstitle{Shedding Coefficient}
\autoref{fig:exp_coefficient} complements the above results with an experiment 
on the shedding coefficient for the individual datasets, for a latency 
threshold of $\theta=10ms$ (similar trends are observed for other $\theta$ 
values). Again, load shedding based on coefficient modelling yields the best 
results. In comparison, the stream elements shed by our method 
do not introduce as many false negatives and false positives as shedding with 
the baseline techniques. The largest difference is observed for the 
\emph{Politics} dataset, which 
features rumours with a complex propagation structure.

\begin{figure}[!h]
  \centering
  \begin{subfigure}{.49\linewidth}
    \centering
    \includegraphics[width=1.0\linewidth]{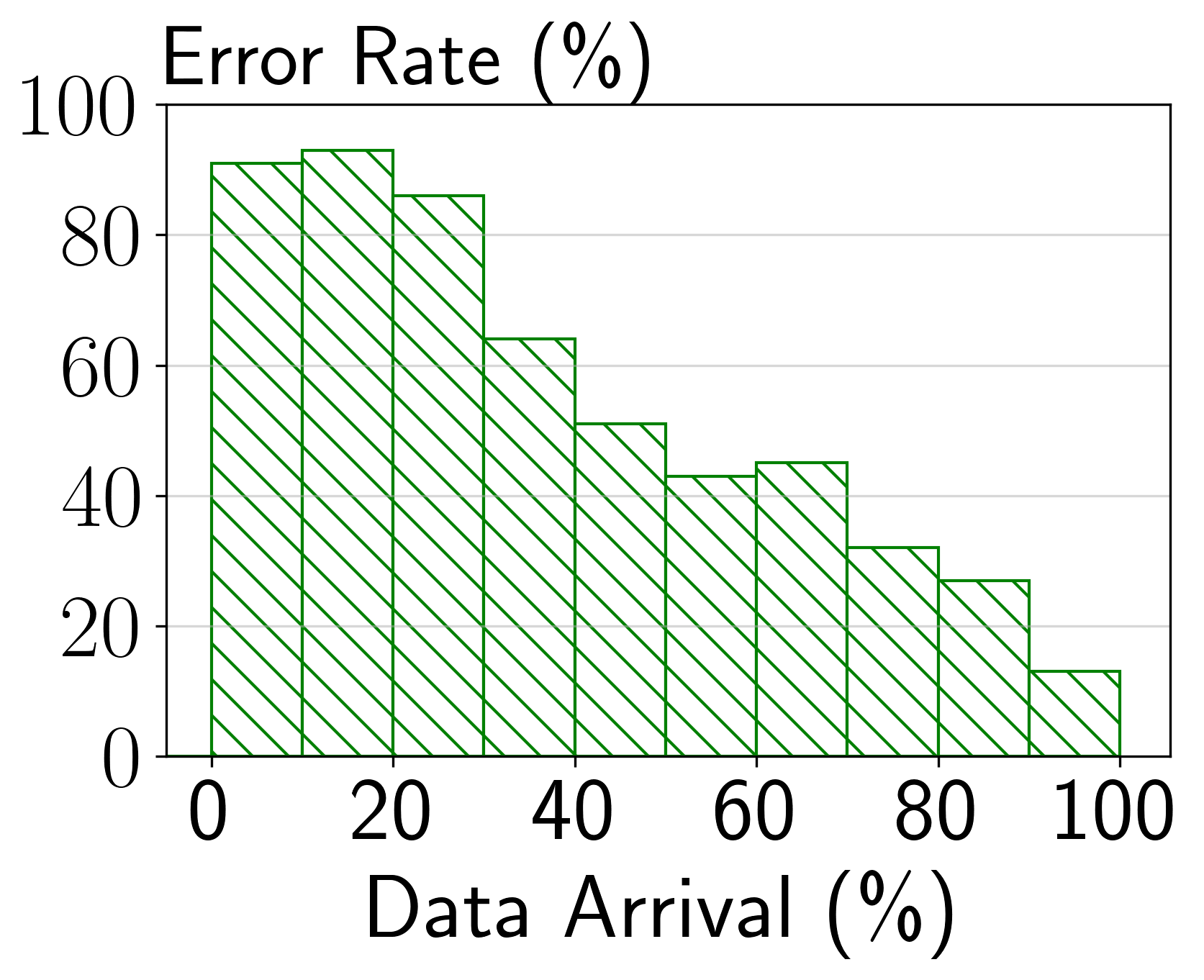}
\vspace{-15pt}
    \caption{Our approach}
  \end{subfigure}
  \begin{subfigure}{.49\linewidth}
    \centering
    \includegraphics[width=1.0\linewidth]{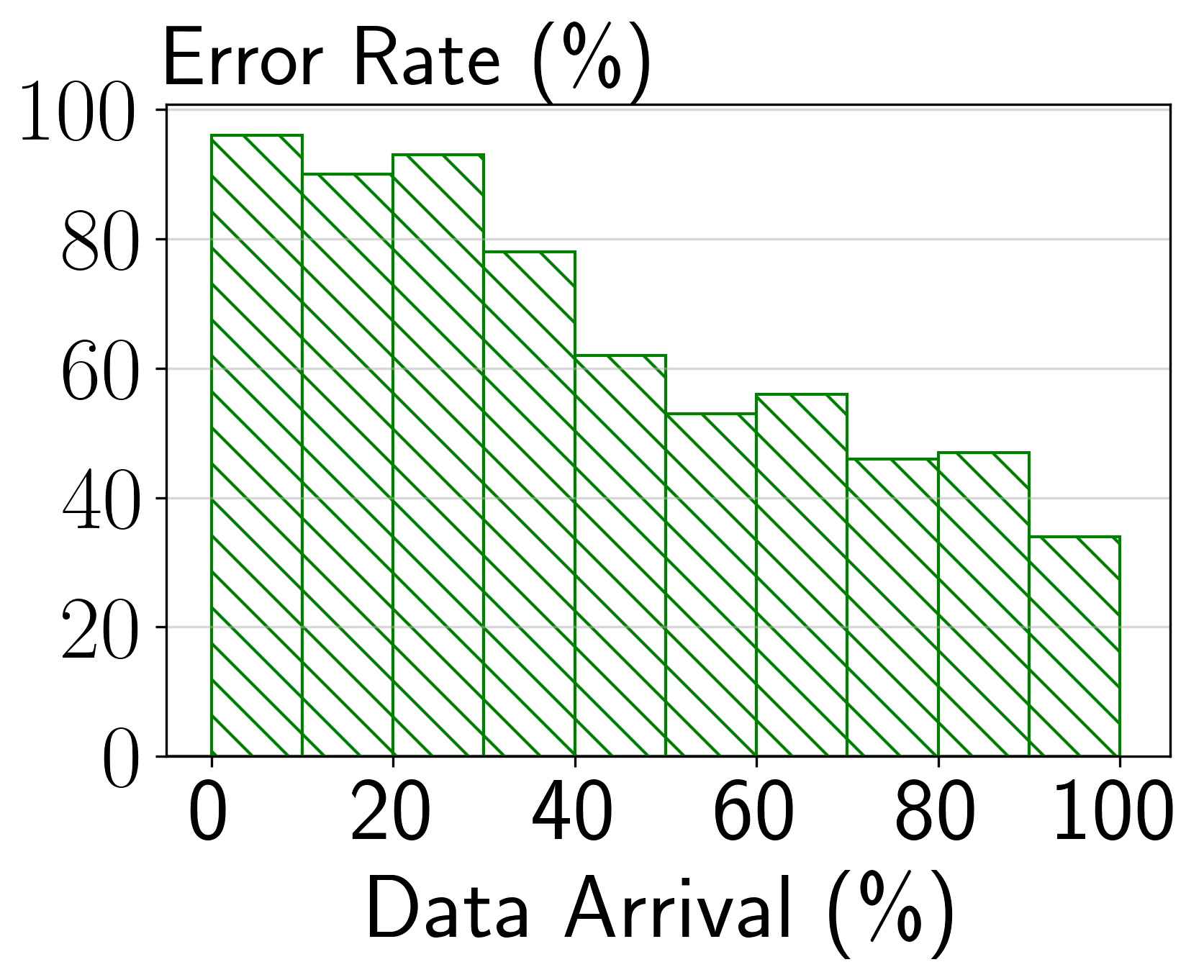}
\vspace{-15pt}
    \caption{\emph{Weighted} baseline}
  \end{subfigure}
\vspace{-5pt}
  \caption{Error rate with different shedding strategies (Politics dataset).}
  \label{fig:exp_error_politics}
\vspace{-10pt}
\end{figure}

\begin{figure}[!h]
  \centering
  \begin{subfigure}{.49\linewidth}
    \centering
    \includegraphics[width=1.0\linewidth]{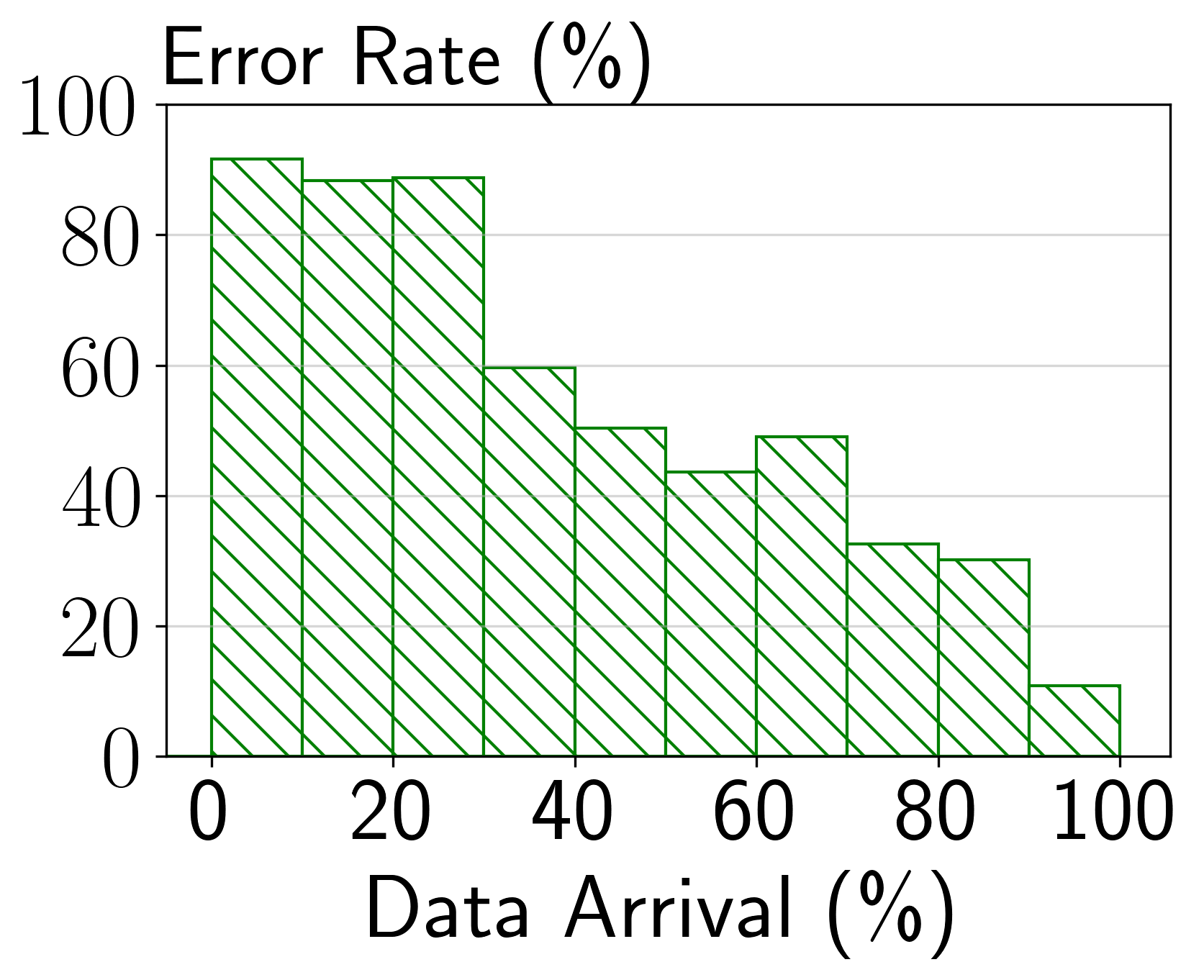}
\vspace{-15pt}
    \caption{Our approach}
  \end{subfigure}
  \begin{subfigure}{.49\linewidth}
    \centering
    \includegraphics[width=1.0\linewidth]{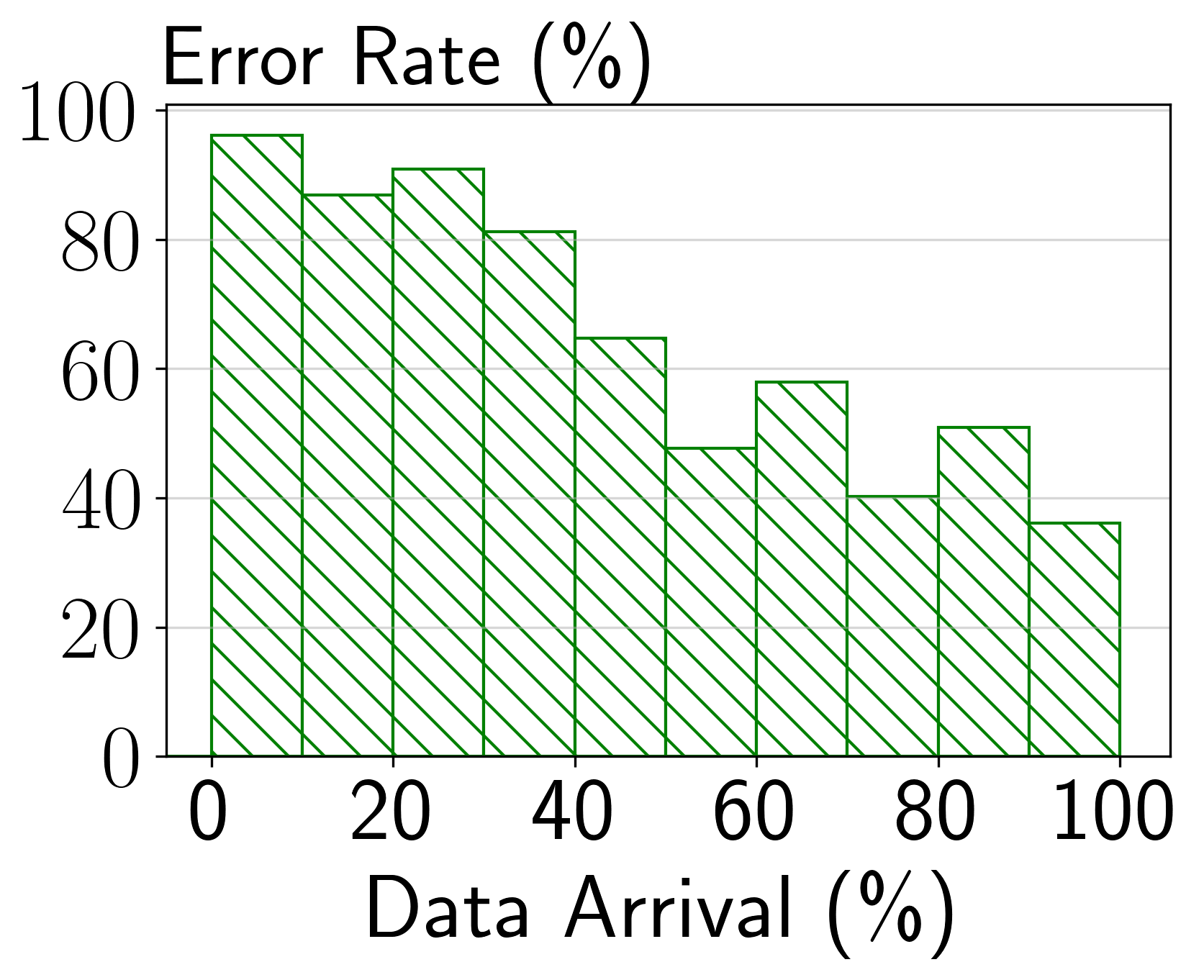}
\vspace{-15pt}
    \caption{\emph{Weighted} baseline}
  \end{subfigure}
\vspace{-5pt}
  \caption{Error rate with different shedding strategies (Crime dataset).}
  \label{fig:exp_error_crime}
\vspace{-10pt}
\end{figure}

\begin{figure}[!h]
  \centering
  \begin{subfigure}{.49\linewidth}
    \centering
    \includegraphics[width=1.0\linewidth]{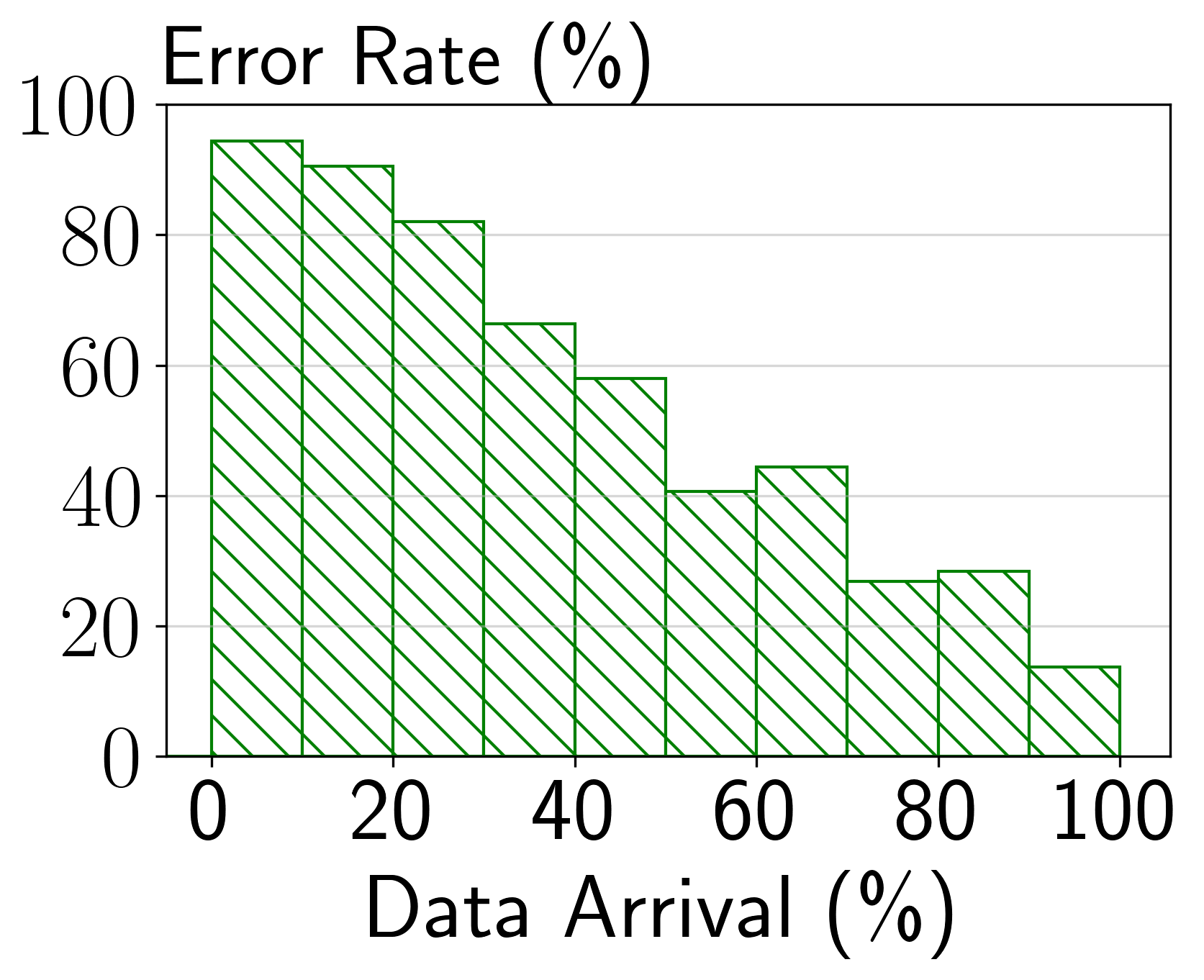}
\vspace{-15pt}
    \caption{Our approach}
  \end{subfigure}
  \begin{subfigure}{.49\linewidth}
    \centering
    \includegraphics[width=1.0\linewidth]{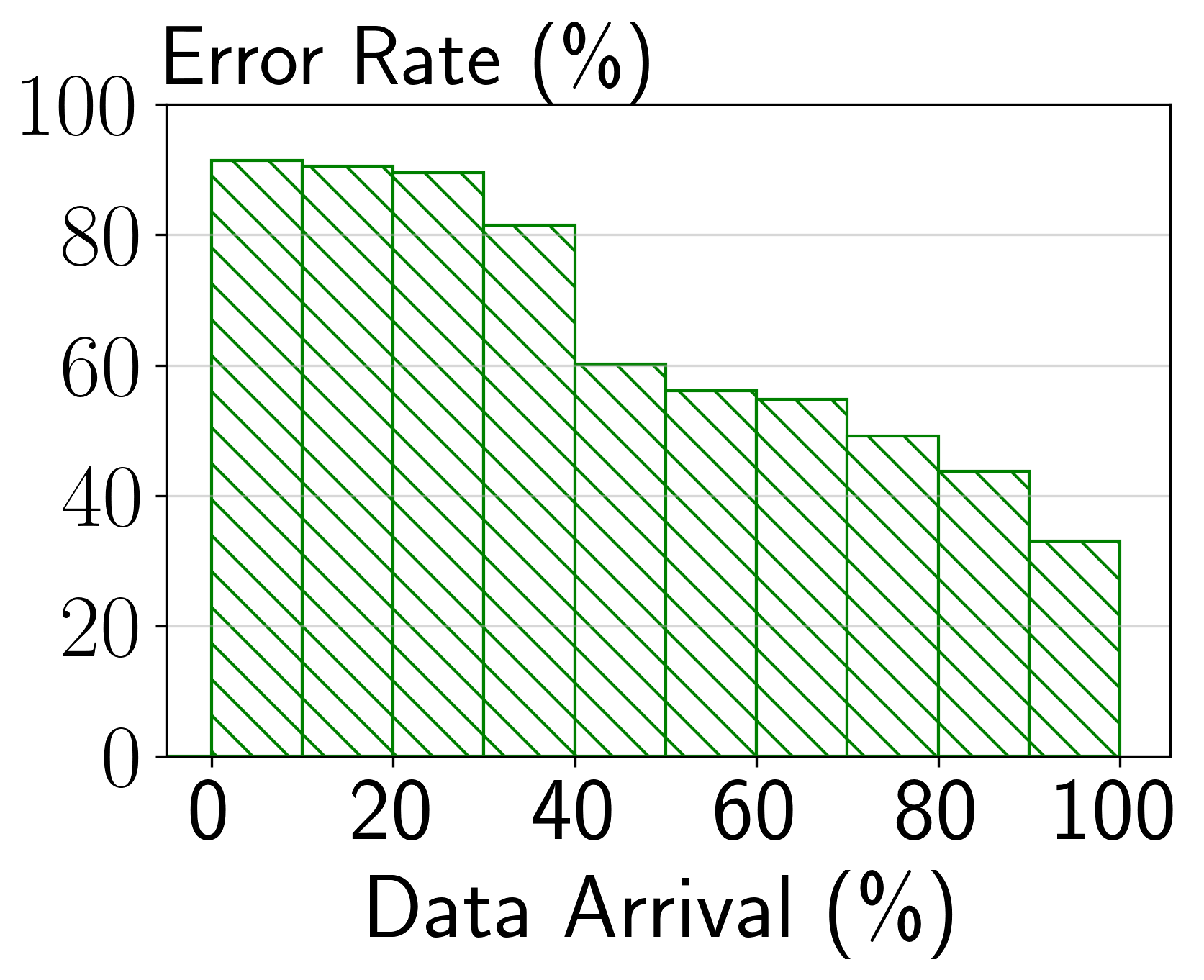}
\vspace{-15pt}
    \caption{\emph{Weighted} baseline}
  \end{subfigure}
\vspace{-5pt}
  \caption{Error rate with different shedding strategies (SciTech dataset).}
  \label{fig:exp_error_scitech}
\vspace{-10pt}
\end{figure}

\sstitle{Error Rate}
We study the effects of data arrival (in \% on the 
x-axis) on the error rate induced by load shedding. 
\autoref{fig:exp_error_politics}, \autoref{fig:exp_error_crime}, and 
\autoref{fig:exp_error_scitech} illustrates the results for the \emph{Politics} 
dataset, the \emph{Crime} dataset, and the \emph{SciTech} dataset, 
respectively.
The error rate is high initially as the coefficient model is not yet 
stable. 
Later, the error rate is reduced significantly, which 
emphasises the correctness of our coefficient model.

\subsection{Effectiveness of Rumour Detection}
\label{sec:exp_accuracy_detect}

Having evaluated the effectiveness of different shedding strategies, we now 
turn to the evaluation of the approaches to rumour detection. That is, we 
compare our approach for streaming rumour detection against the ground truth as 
well as the state-of-the-art techniques mentioned in \autoref{sec:exp_setup}.
In any case, rumour detection is conducted with our strategy for load shedding, 
since the superiority of this strategy has been demonstrated in the previous 
experiments.

\begin{figure}[!h]
  \centering
  \begin{subfigure}{.49\linewidth}
    \centering
    \includegraphics[width=1.\linewidth]{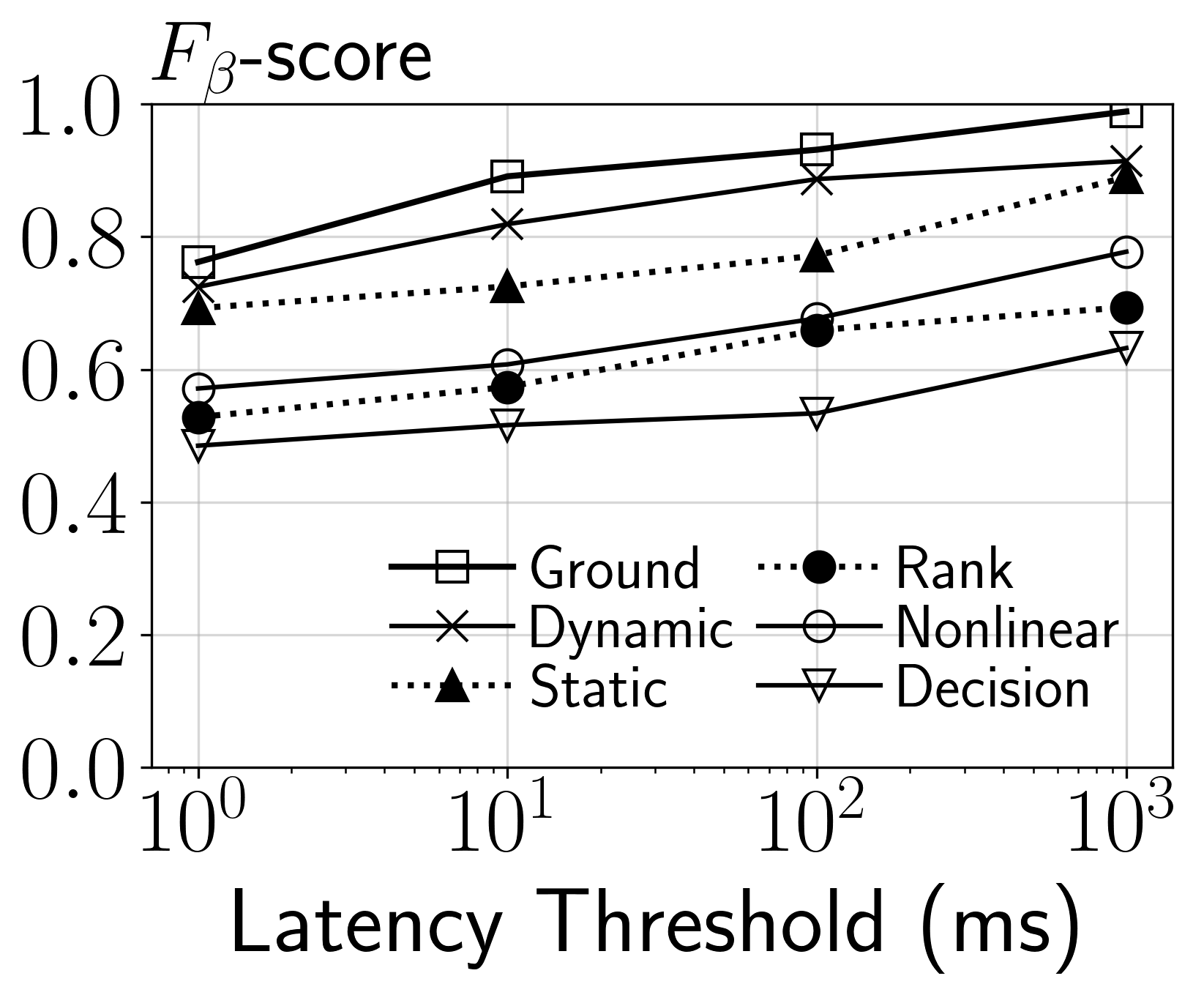}
    \caption{Politics dataset}
  \label{fig:exp_accuracy_detector_politics}
  \end{subfigure}
  \begin{subfigure}{.49\linewidth}
    \centering
    \includegraphics[width=1\linewidth]{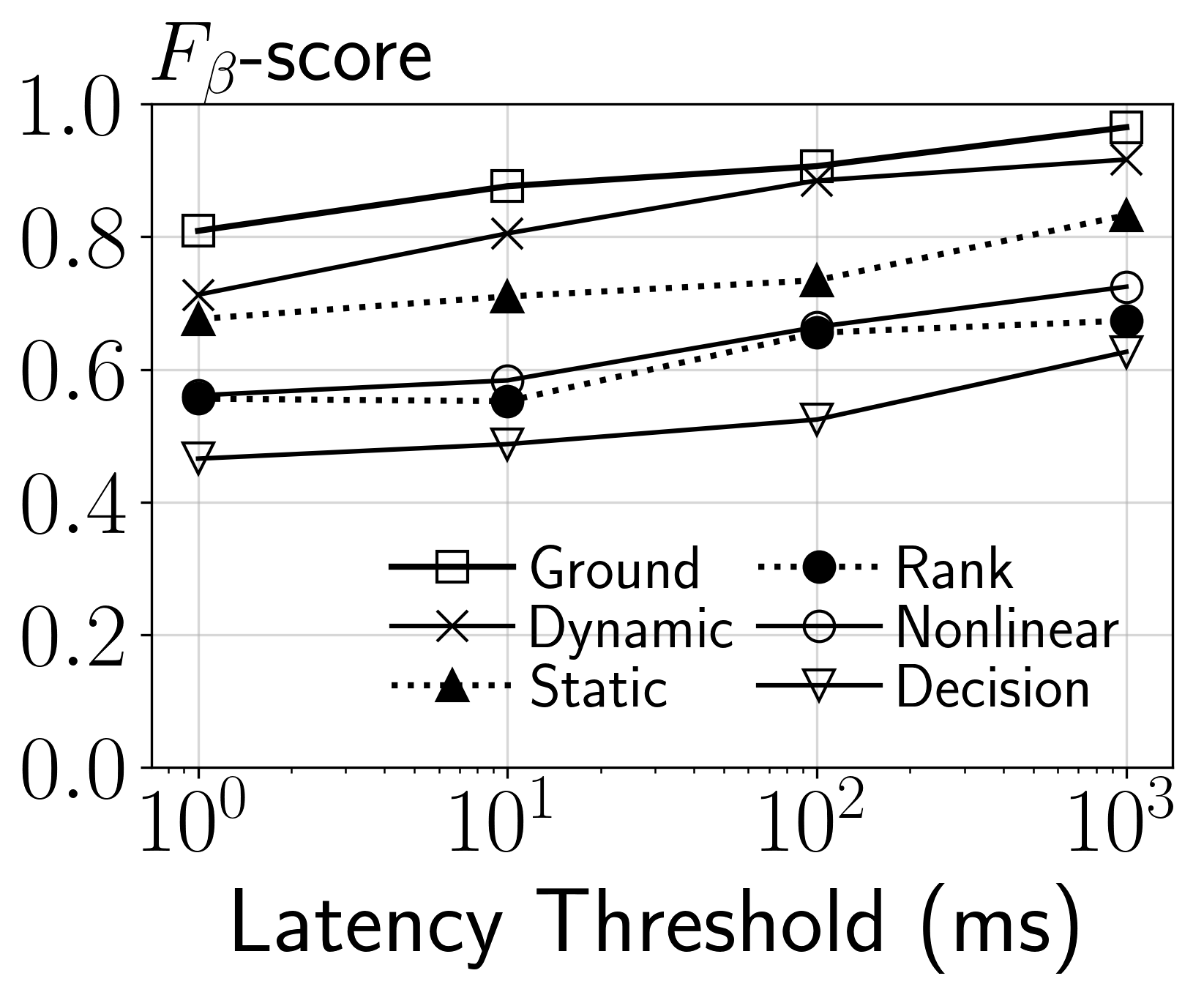}
    \caption{Crime dataset}
  \label{fig:exp_accuracy_detector_crime}
  \end{subfigure}
  \begin{subfigure}{.49\linewidth}
    \centering
    \includegraphics[width=1\linewidth]{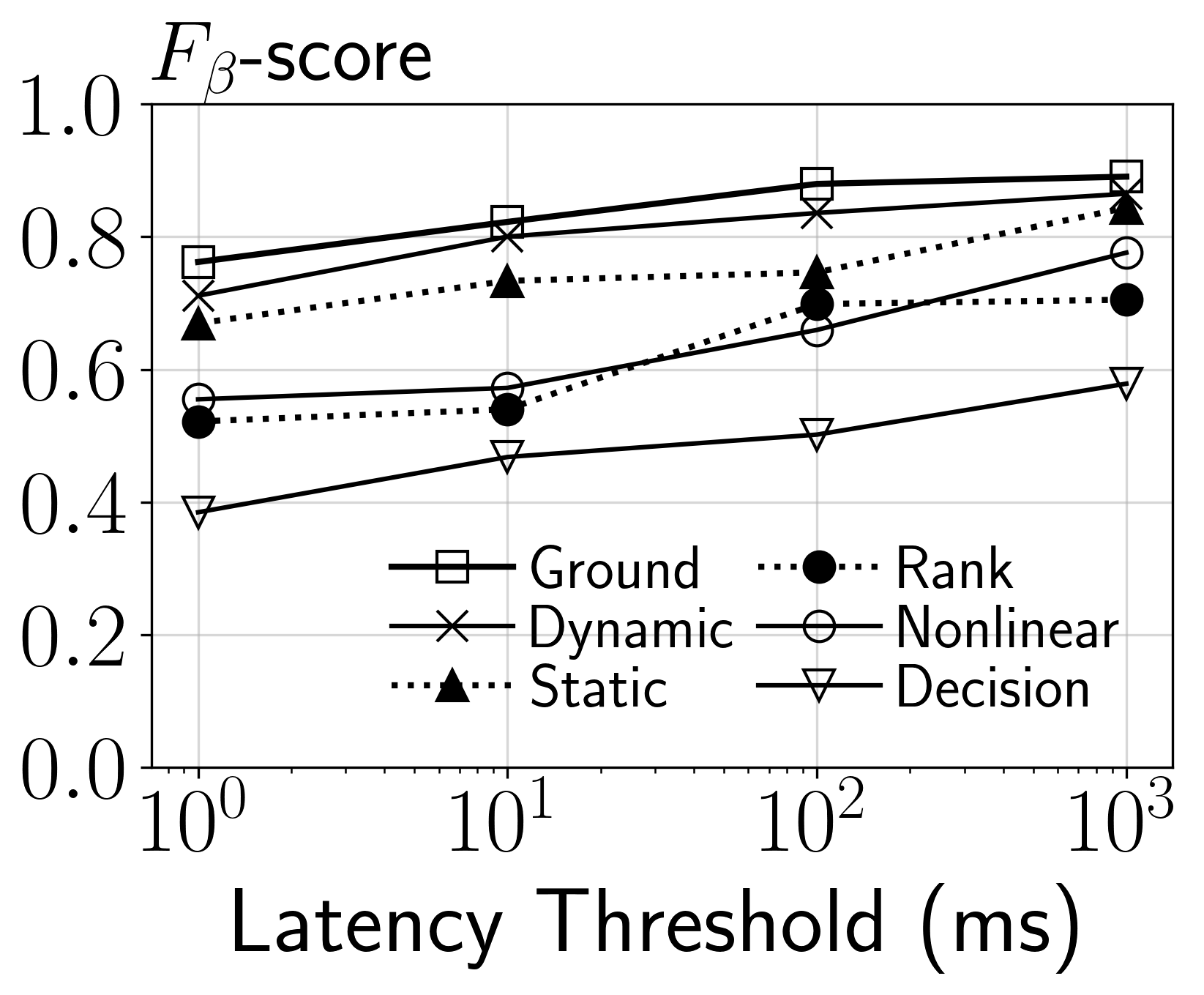}
    \caption{SciTech dataset}
  \label{fig:exp_accuracy_detector_scitech}
  \end{subfigure}
  \caption{Accuracy of rumour detection.}
  \label{fig:exp_accuracy_detector}
\vspace{-1.5em}
\end{figure}

\sstitle{Accuracy}
\autoref{fig:exp_accuracy_detector} compares the performance of different 
rumour detection techniques under various latency constraints, for each 
dataset. In general, streaming rumour detection as proposed in this paper 
(\emph{dynamic}) 
outperforms the baseline techniques, and is closest to the actual ground truth. 
 

We further note that the \emph{static} version of anomaly-based 
rumour detection yields the next best results. Moreover, the results, in 
general, become better with higher latency thresholds. This may be explained by 
less aggressive load shedding.

\begin{figure}[!h]
\centering
    \includegraphics[width=0.5\linewidth]{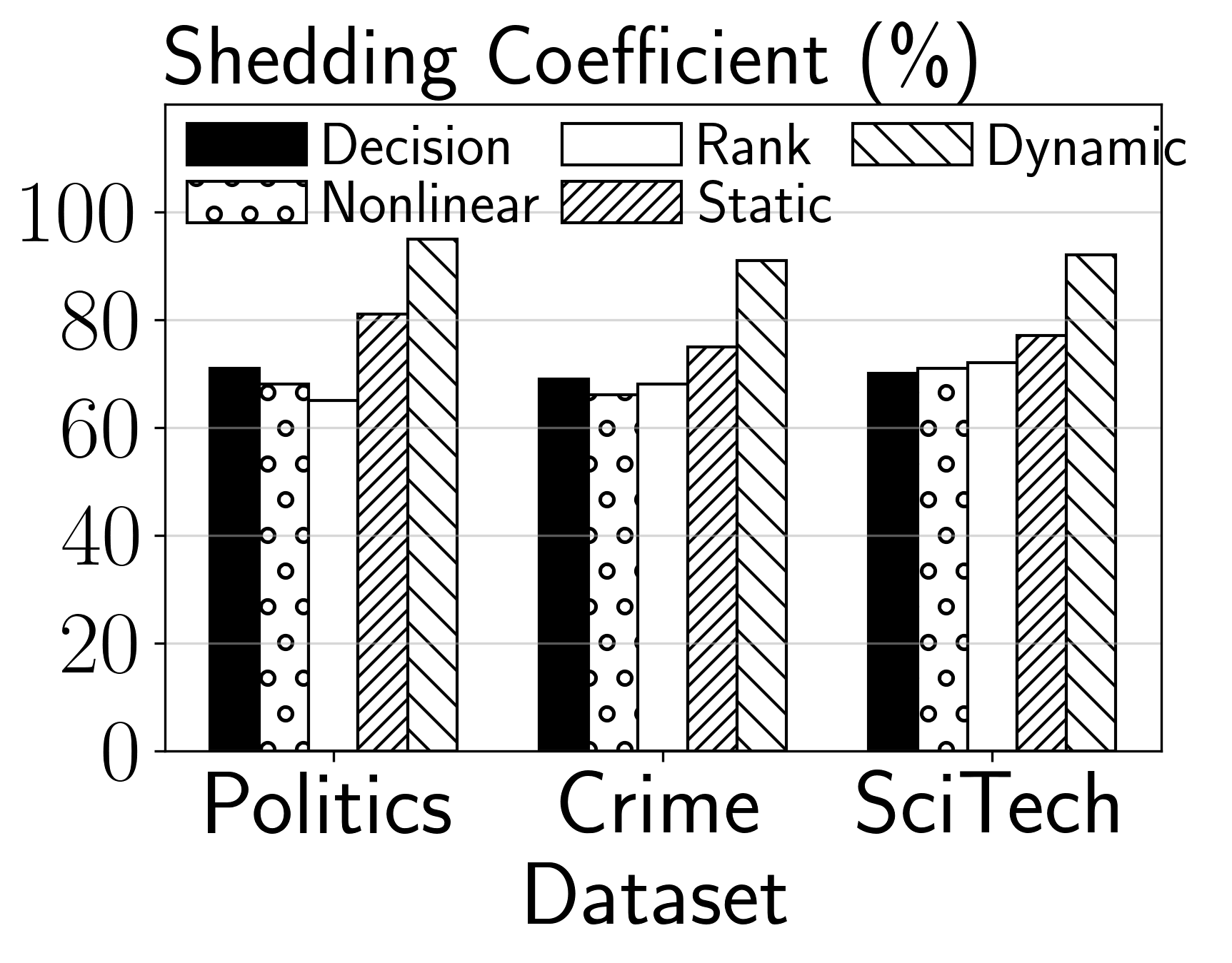}
    \vspace{-5pt}
    \caption{Coefficient of rumour detection.}
  \label{fig:exp_coefficient_detector}
\end{figure}

\sstitle{Shedding Coefficient}
Similar to the previous experiment, we compare the shedding coefficient for 
different datasets for various rumour detection strategies. \autoref{fig:exp_coefficient_detector} reports the results.

In general, rumour detection with higher accuracy leads to larger shedding 
coefficients. This is explained by the fact that accurate detection leads 
to stable coefficient modelling, which is the basis for our shedding strategy.

\begin{figure}[!h]
  \centering
  \begin{subfigure}{.49\linewidth}
    \centering
    \includegraphics[width=1.0\linewidth]{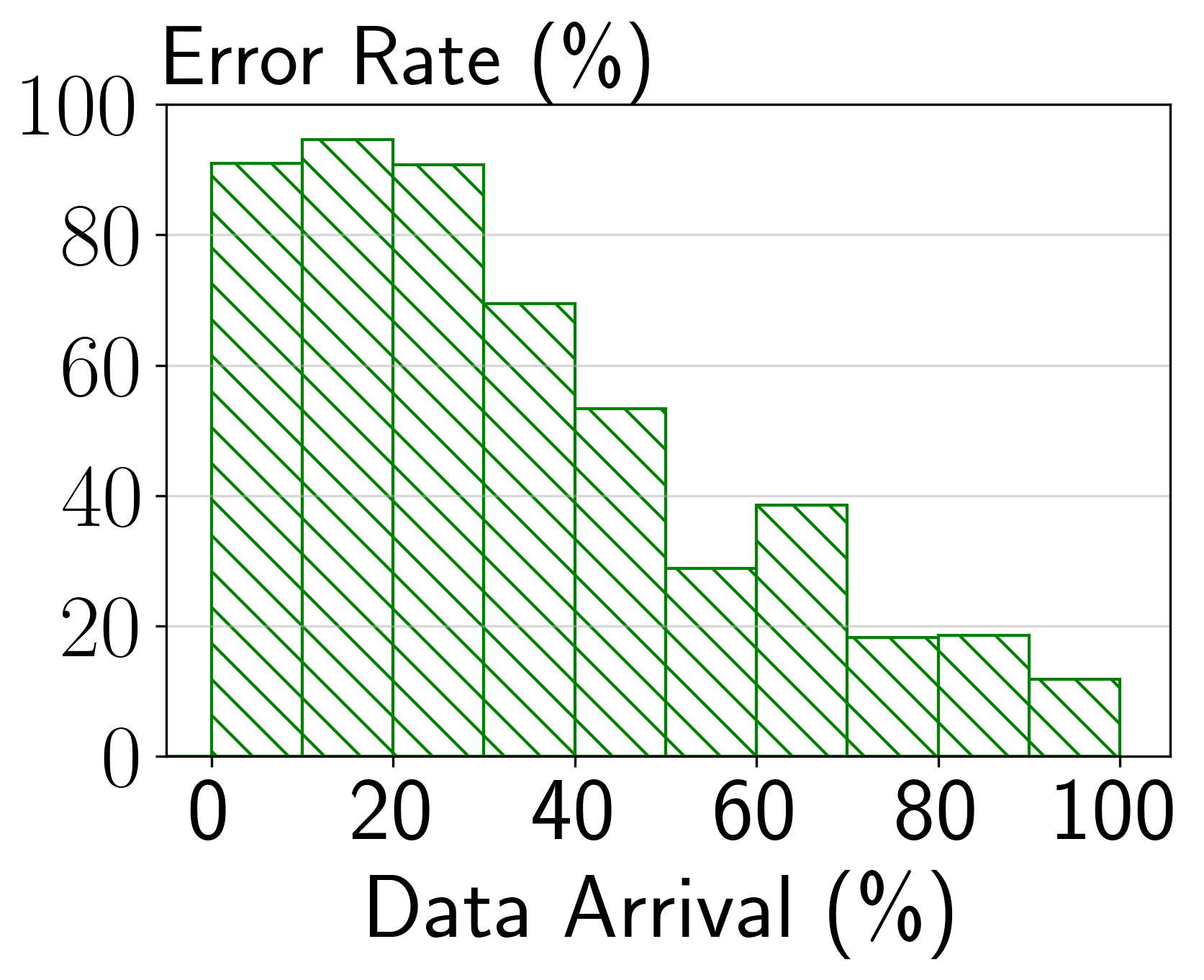}
\vspace{-15pt}
    \caption{\emph{Dynamic} detector}
  \end{subfigure}
  \begin{subfigure}{.49\linewidth}
    \centering
    \includegraphics[width=1.0\linewidth]{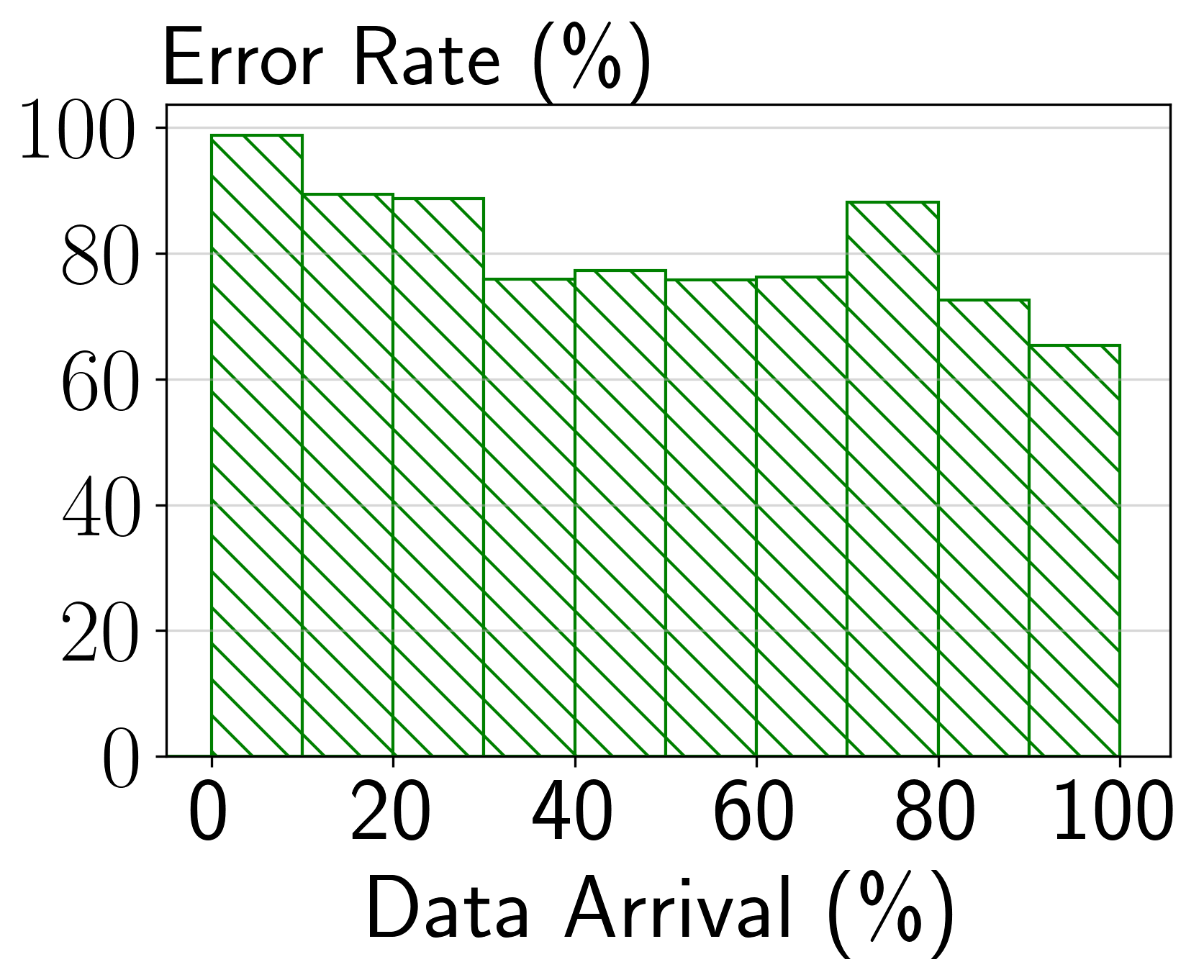}
\vspace{-15pt}
    \caption{\emph{Static} detector}
  \end{subfigure}
\vspace{-5pt}
  \caption{Error rate with different rumour detectors (Politics dataset).}
  \label{fig:exp_error_detector_politics}
\vspace{-10pt}
\end{figure}

\begin{figure}[!h]
  \centering
  \begin{subfigure}{.49\linewidth}
    \centering
    \includegraphics[width=1.0\linewidth]{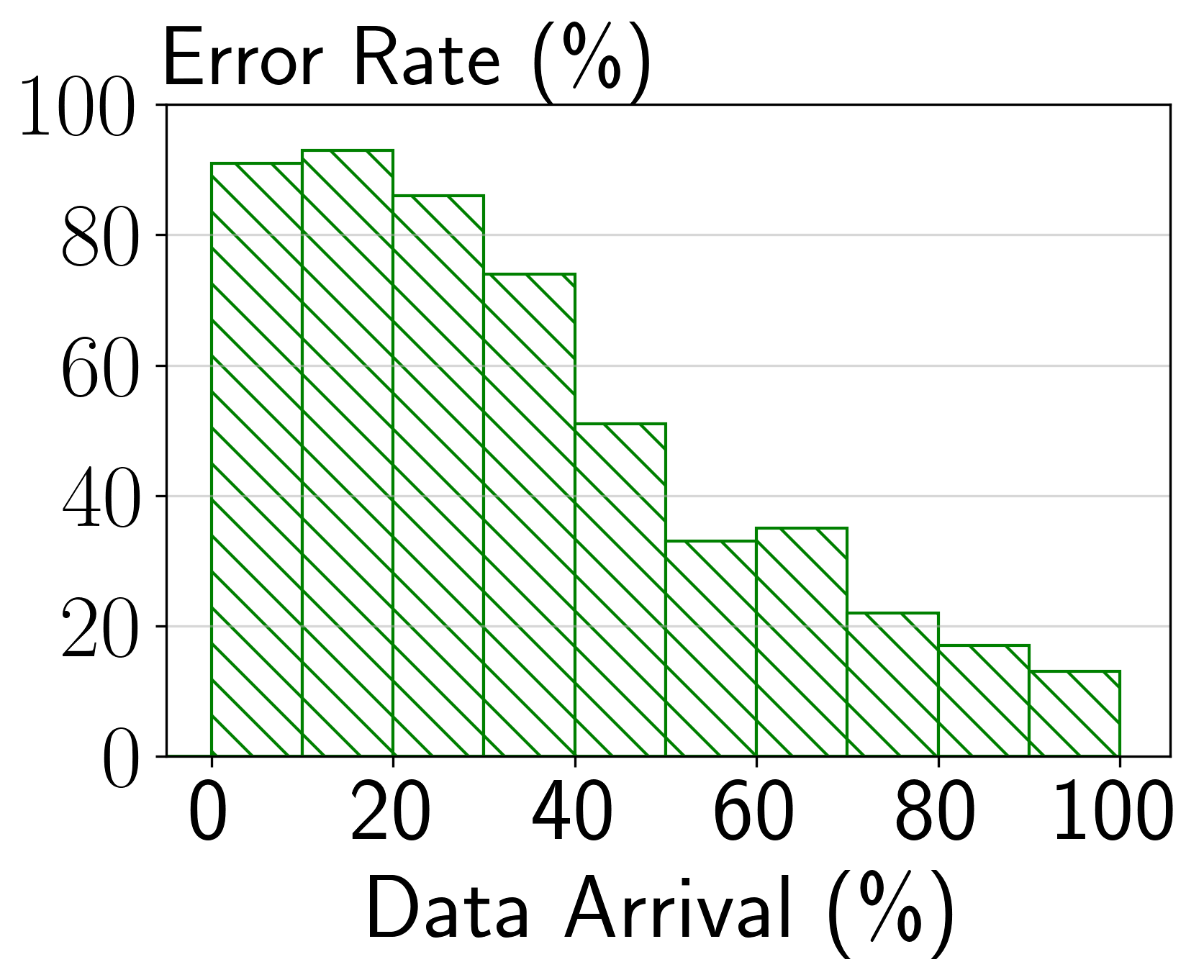}
\vspace{-15pt}
    \caption{\emph{Dynamic} detector}
  \end{subfigure}
  \begin{subfigure}{.49\linewidth}
    \centering
    \includegraphics[width=1.0\linewidth]{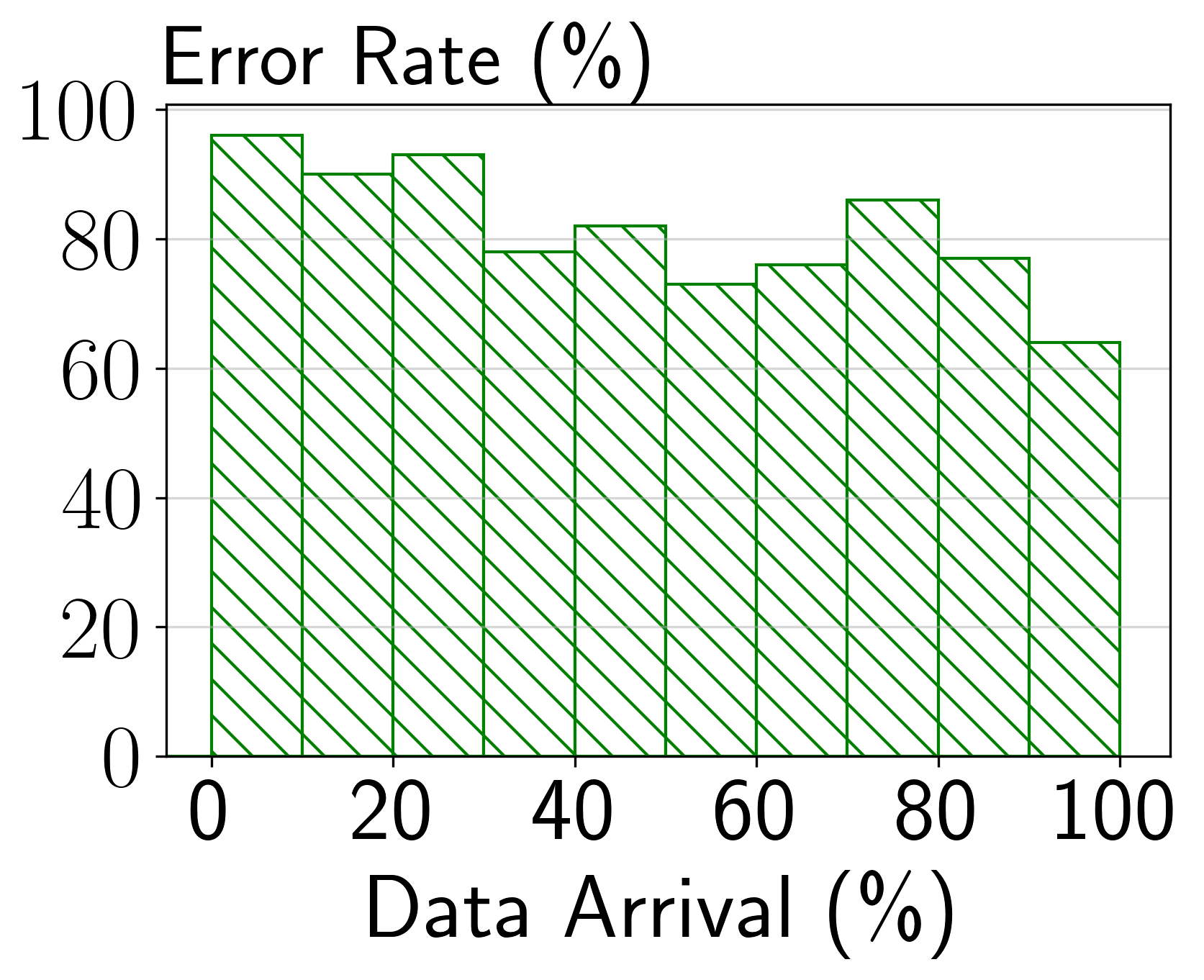}
\vspace{-15pt}
    \caption{\emph{Static} detector}
  \end{subfigure}
\vspace{-5pt}
  \caption{Error rate with different rumour detectors (Crime dataset).}
  \label{fig:exp_error_detector_crime}
\vspace{-10pt}
\end{figure}

\begin{figure}[!h]
  \centering
  \begin{subfigure}{.49\linewidth}
    \centering
    \includegraphics[width=1.0\linewidth]{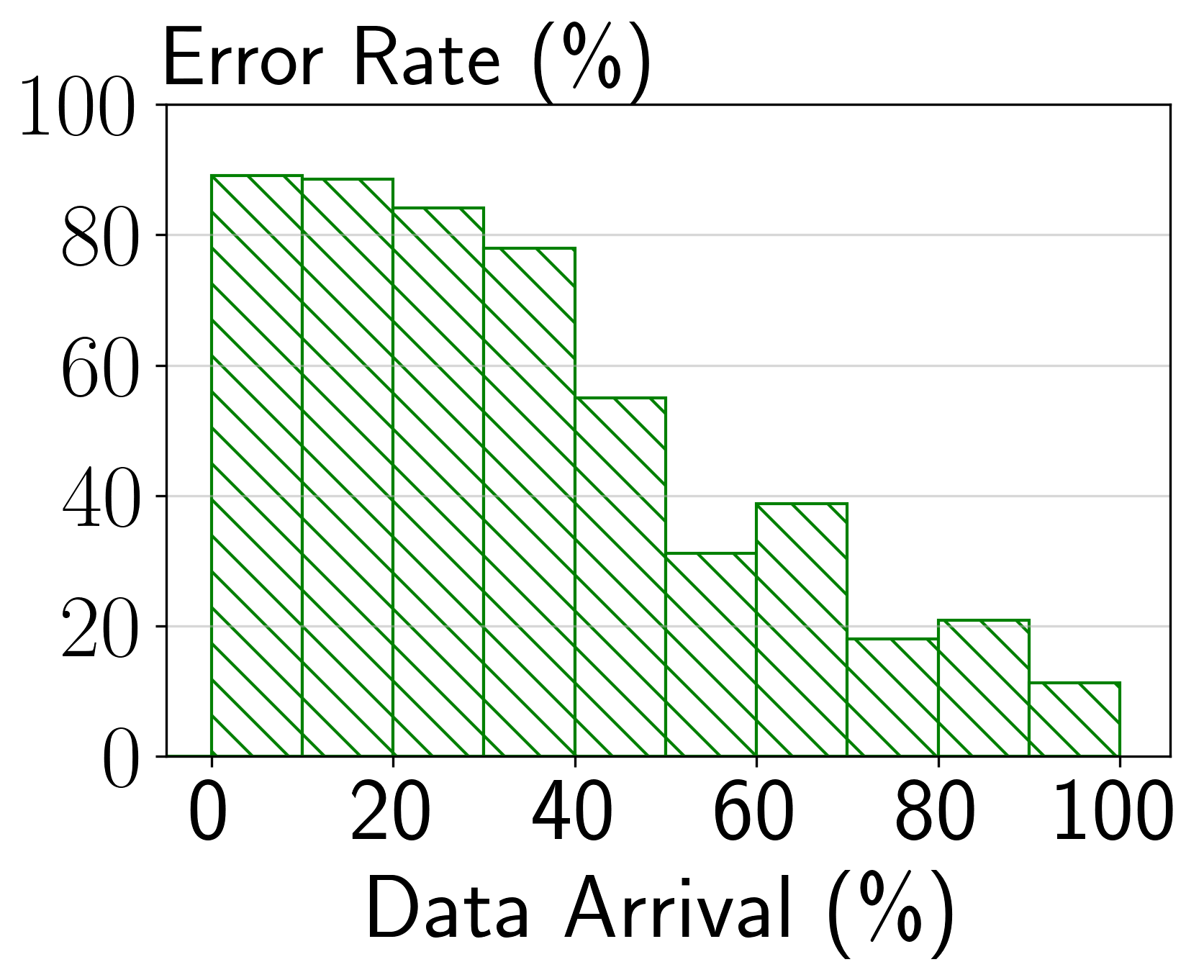}
\vspace{-15pt}
    \caption{\emph{Dynamic} detector}
  \end{subfigure}
  \begin{subfigure}{.49\linewidth}
    \centering
    \includegraphics[width=1.0\linewidth]{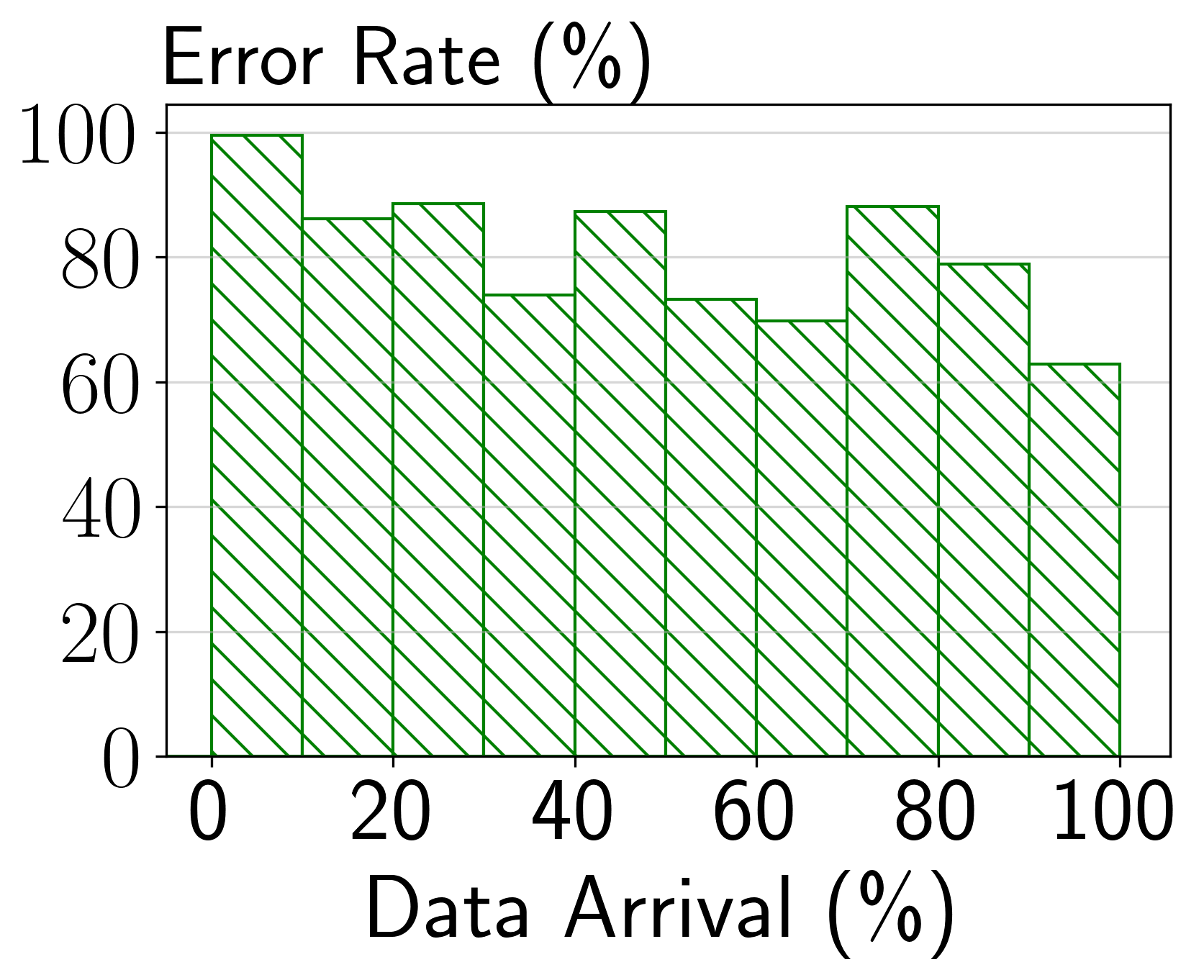}
\vspace{-15pt}
    \caption{\emph{Static} detector}
  \end{subfigure}
\vspace{-5pt}
  \caption{Error rate with different rumour detectors (SciTech dataset).}
  \label{fig:exp_error_detector_scitech}
\vspace{-10pt}
\end{figure}

\sstitle{Error Rate}
We compare the shedding error when using our 
strategy for streaming rumour detection (\emph{dynamic}) and against the 
next-best approach, i.e., \emph{static} rumour detection based on anomalies. 
\autoref{fig:exp_error_detector_politics}, 
\autoref{fig:exp_error_detector_crime}, and 
\autoref{fig:exp_error_detector_scitech} illustrates the results for the 
\emph{Politics} 
dataset, the \emph{Crime} dataset, and the \emph{SciTech} dataset, 
respectively.
Here, we observe that our technique converges 
relatively quick to low errors, whereas the error stays at high values with the 
\emph{static} strategy. This reason for this observation is the fact that the 
\emph{static} strategy does not give a latency guarantee. Hence, the system may 
shed many stream elements that are important to reach stable 
coefficient modelling. Results with the other rumour detection strategies show a similar trend.

\subsection{Efficiency of Shedding Strategies}
\label{sec:exp_time_shedding}

Next, we assess the efficiency of the proposed approach, starting with an 
analysis of the shedding strategies. 

\begin{figure}[!h]
\centering
    \includegraphics[width=0.5\linewidth]{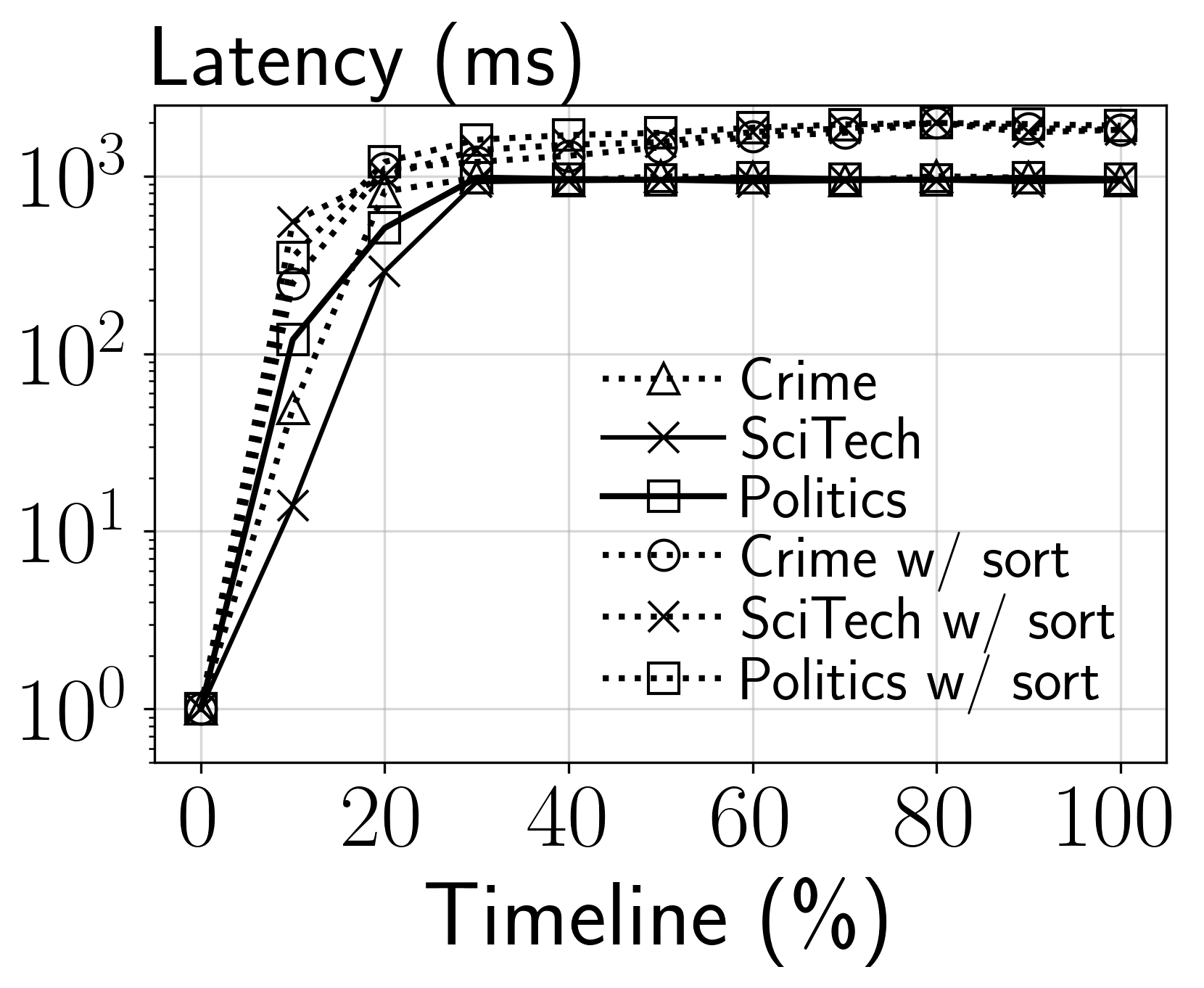}
\vspace{-5pt}
    \caption{Latency of shedding strategies.}
  \label{fig:exp_latency}
  \end{figure}

\sstitle{Latency}
\autoref{fig:exp_latency} depicts the latency observed during processing 
relative to the elapsed time (in percentage of the covered time period), which 
is calculated by the timestamp of the incoming data 
($\theta = 10^3ms$). 
For all three datasets, the latency stays constant after reaching a plateau of 
around 1$s$. As another reference point, we include a \emph{sort} version of 
our shedding method in \autoref{sec:shedding}, which sorts the window by 
coefficients and sheds the $k$ elements with the lowest coefficient values. 
This variant incurs higher latency, which points to the importance of 
thresholding as employed in our approach.  

\begin{figure}[!h]
\centering
    \includegraphics[width=0.5\linewidth]{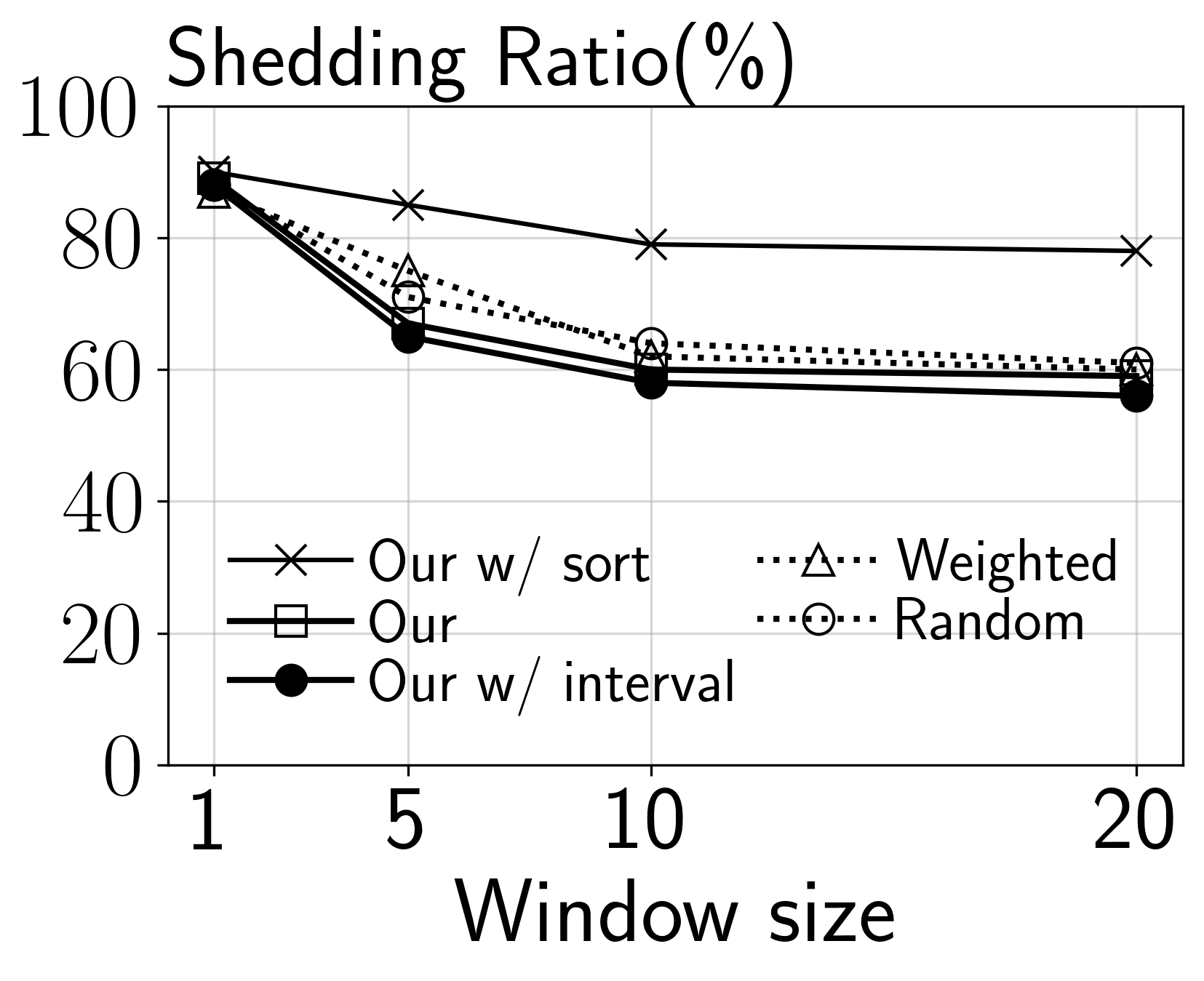}
\vspace{-5pt}
    \caption{Shedding granularity.}
  \label{fig:exp_shedding}
  \end{figure}

\sstitle{Shedding Ratio}
\autoref{fig:exp_shedding} presents the percentage of shed data (smaller is 
better), while varying the window size ($\theta = 1ms$). The 
\emph{sort} version of the load shedding procedure performs worst, due to the 
sorting overhead and the need to shed more data in the next window. Our actual 
method has similar performance as the \emph{weighted} and 
\emph{random} baselines since they use the same decision on when and how much 
data to shed. Our extension that includes a variable shedding \emph{interval} 
(\autoref{sec:extension}) performs 
best, as shedding is conducted at a more fine-grained level.

\subsection{Efficiency of Rumour Detection}
\label{sec:exp_time_detect}

We complement the above results with experiments on the latency and shedding 
ratio of different strategies for rumour detection. As above, we combine these 
techniques with our approach to load shedding.

\begin{figure}[!h]
    \centering
    \includegraphics[width=0.5\linewidth]{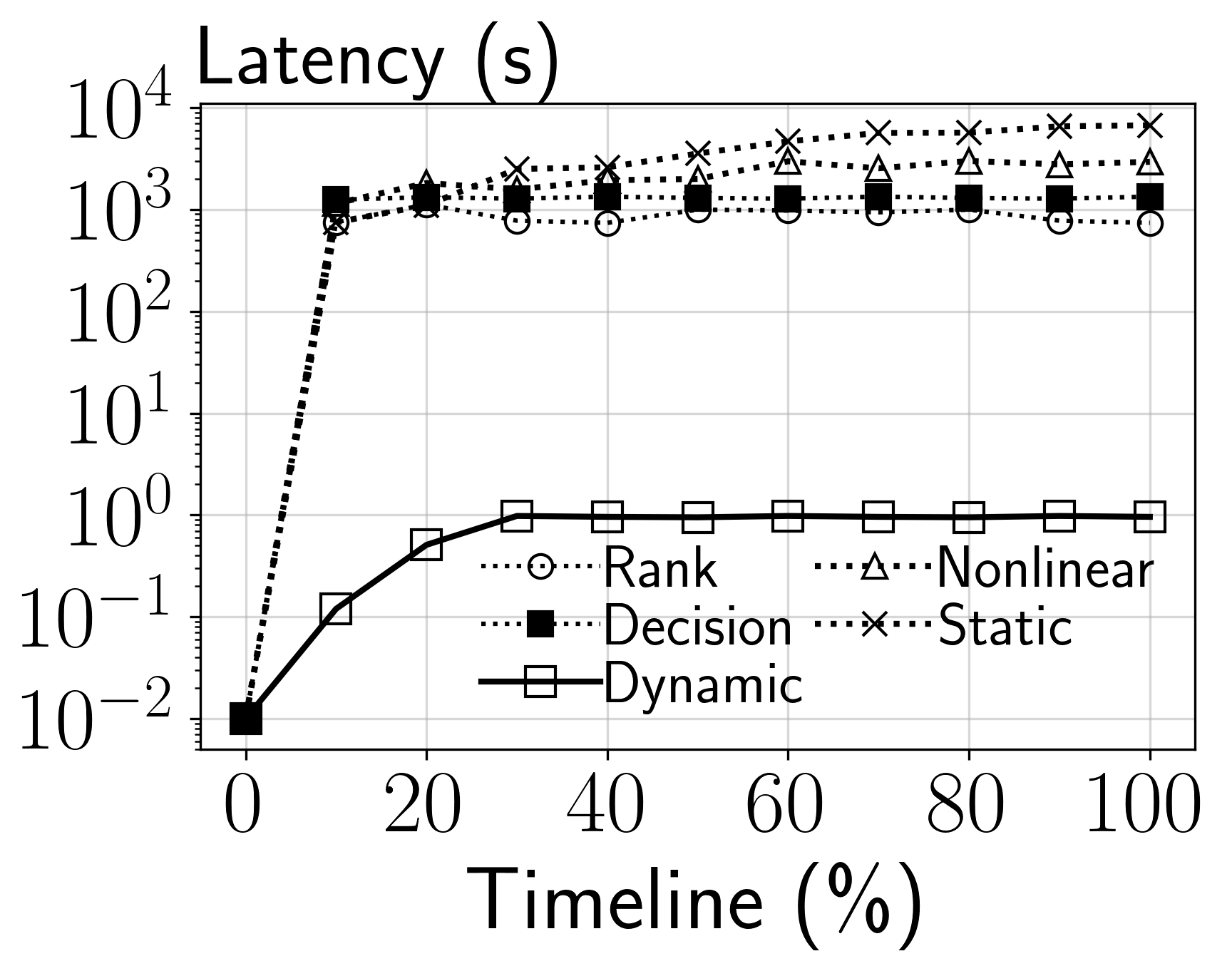}
\vspace{-5pt}
    \caption{Latency of rumour detection.}
  \label{fig:exp_latency_detector}
  \end{figure}

\sstitle{Latency}
\autoref{fig:exp_latency_detector} illustrates the latency observed for 
different rumour detection strategies, under a latency threshold of $\theta = 
1s$. Again, the latency is considered relative to the elapsed time. It can be 
seen that our proposed \emph{dynamic} strategy to rumour detection 
guarantees the latency constraint, while the other strategies violate it. 

\begin{figure}[!h]
    \centering
    \includegraphics[width=0.5\linewidth]{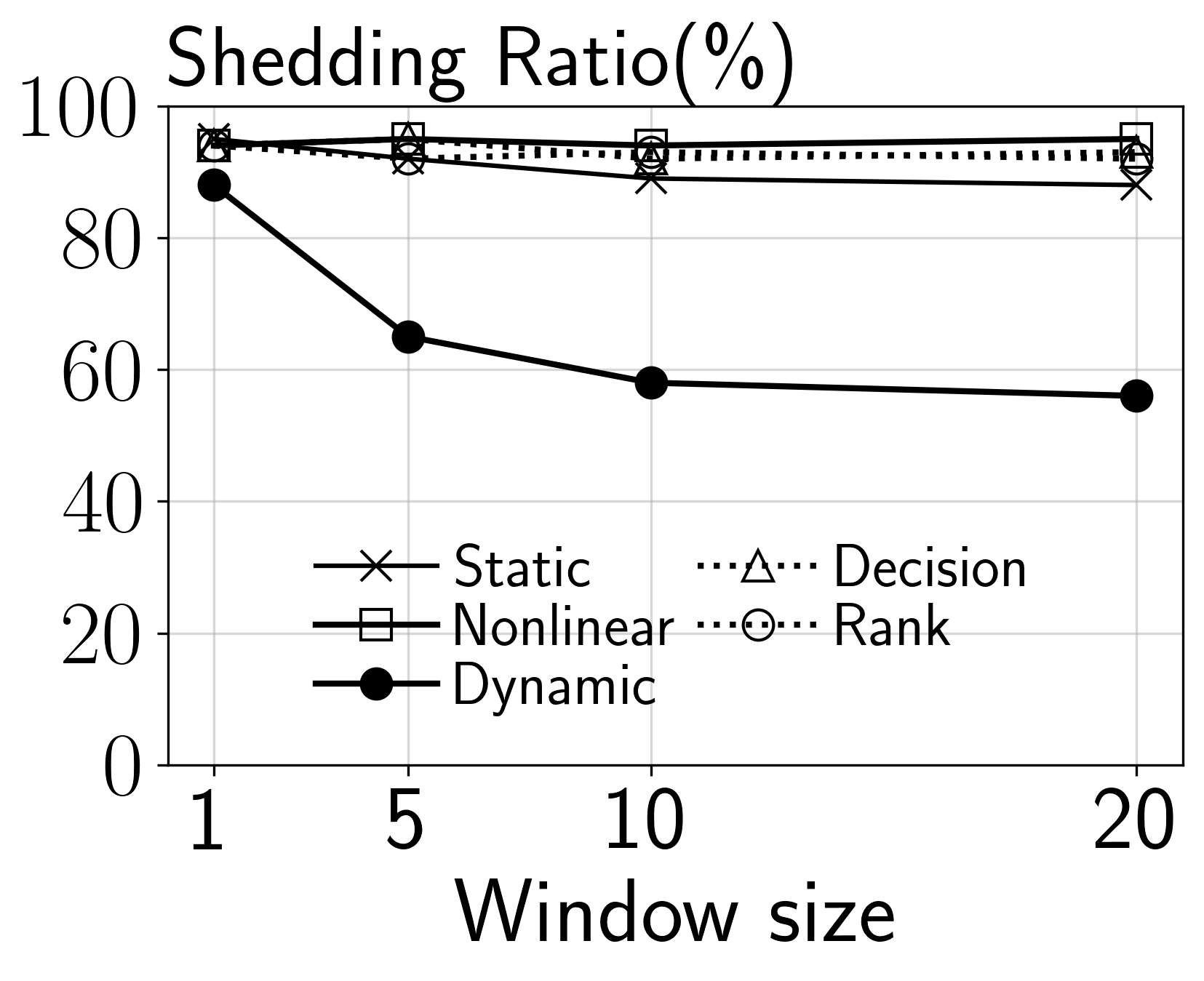}
\vspace{-5pt}
    \caption{Shedding ratio of rumour detection.}
  \label{fig:exp_shedding_detector}
  \end{figure}

\sstitle{Shedding Ratio}
\autoref{fig:exp_shedding_detector} depicts the percentage of shed data 
(smaller is better) for different rumour detection strategies. We vary the 
window size, while fixing the latency threshold to $\theta = 1s$. In general, 
our proposed \emph{dynamic} strategy outperforms the baseline techniques. This 
is because the other strategies violate the latency threshold and, hence, 
increase the amount of data that needs to be shed. 

\begin{figure}[!h]
    \centering
    \includegraphics[width=0.5\linewidth]{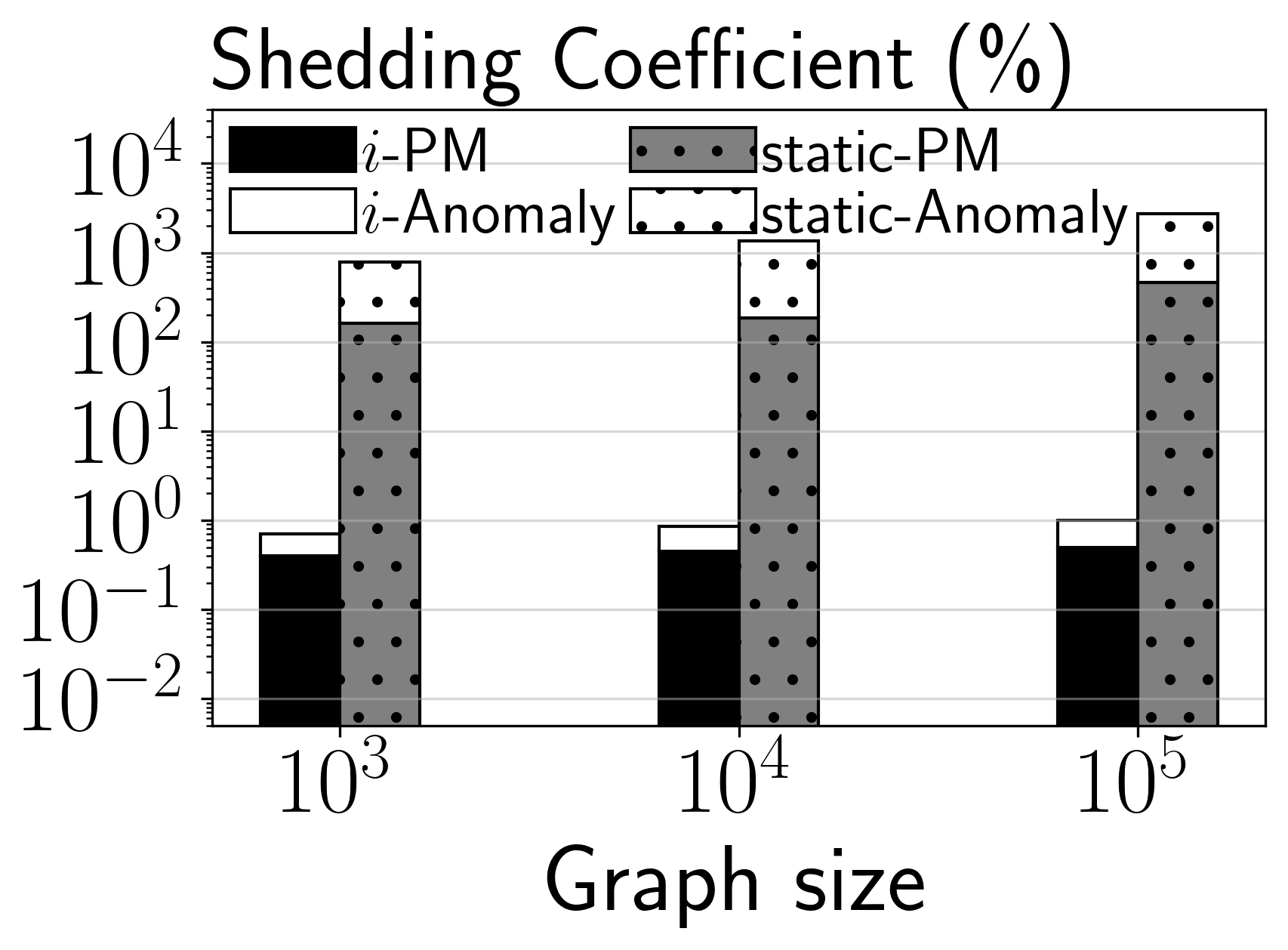}
\vspace{-5pt}
    \caption{Runtime of streaming rumour detection.}
  \label{fig:exp_runtime}
  \end{figure}

  \sstitle{Runtime}
\autoref{fig:exp_runtime} reports the runtime of streaming pattern 
 matching (\autoref{sec:incremental_matching}, namely $i$-PM) and streaming 
 anomaly computation (\autoref{sec:incremental_anomaly}, namely $i$-Anomaly) 
 against their offline counterparts for different graph sizes. We use 
 GraphGen,\footnote{\url{https://cse.hkust.edu.hk/graphgen/}} to generate 
 social graphs of different sizes and features. 
 Here, \emph{static-PM} is the offline counterpart of $i$-PM that uses 
 SPMiner~\cite{SPMiner2020}, a state-of-the-art method to 
 find matches of subgraph patterns based on subgraph isomorphism. Moreover, 
 \emph{static-Anomaly} denotes the offline 
 counterpart of $i$-Anomaly, which computes anomaly scores from raw historical 
 data~\cite{tam2019anomaly}.

In general, the proposed streaming mechanisms 
 outperform their offline counterparts from one to two orders of magnitude. 
 The stacked bar chart in \autoref{fig:exp_runtime} further reveals that 
 graph-based rumour detectors~\cite{tam2019anomaly,zhao2015enquiring} that use 
 \emph{static-PM} and \emph{static-Anomaly}, in general, are much 
 slower than the streaming approach presented in this work.

\subsection{Sensitivity Analysis}
\label{sec:exp_sensitivity}

\sstitle{Coefficient distribution, input rate, and pattern size}
\autoref{tab:exp_distribution} illustrates the sensitivity of our approach 
against different data distributions (\emph{normal} and \emph{skewed} 
coefficient distributions), input rates (the delay between stream elements 
follows a \emph{normal} 
or \emph{skewed} distribution), and pattern sizes. 
Here, larger pattern sizes lead to better performance since the coefficient 
statistics are more accurate (their matches are more likely to 
become rumours). Skewed input rates lower the performance since more data is 
shed in case of short bursts. However, skewed coefficient distributions improve 
the results since low-coefficient data are more likely to be shed together 
in a window.

\begin{table}[!h]
\scriptsize
\centering
\caption{$F_{\beta}$-score in different settings ($\theta=1ms$).}
\label{tab:exp_distribution}
\vspace{-7pt}
\begin{tabular}{clccc}
\toprule
coefficient dist. & input rate & $|p|=3$ & $|p|=5$ & $|p|=10$ \\
\midrule
\multirow{2}{*}{\rotatebox{0}{ normal}} & normal & 0.81 & 0.82 & 0.85 \\
 & skewed & 0.71 & 0.73 & 0.78 \\
 \midrule
\multirow{2}{*}{\rotatebox{0}{ skewed}} & normal & 0.84 & 0.83 &
0.86 \\
 & skewed & 0.75 & 0.78 & 0.85 \\
\bottomrule
\end{tabular}
\end{table}

\begin{table}[!h]
\vspace{-5pt}
\scriptsize
\centering
\caption{Shedding coefficient with rumour detectors ($\theta=10ms$)}
\label{tab:exp_detector}
\vspace{-7pt}
\begin{tabular}{clccc}
\toprule
window size & pattern size & $f=50\%$ & $f=75\%$ & $f=100\%$ \\
\midrule
\multirow{2}{*}{\rotatebox{0}{ normal}} & normal & 0.63 & 0.91 & 0.95 \\
 & skewed & 0.61 & 0.90 & 0.92 \\
 \midrule
\multirow{2}{*}{\rotatebox{0}{ skewed}} & normal & 0.59 & 0.89 &
0.93 \\
 & skewed & 0.62 & 0.88 & 0.91 \\
\bottomrule
\end{tabular}
\vspace{-10pt}
\end{table}

\sstitle{Effects of detection accuracy}
\autoref{tab:exp_detector} explores the impact of the 
quality of rumour detection, which is simulated by negating the ground-truth 
with different probabilities ($f=50\%,75\%,100\%$). We consider \emph{normal} 
and \emph{skewed} distributions for the pattern size and the window size, to 
also test our extension for variable window sizes (\autoref{sec:extension}).
Our approach turns out to be robust. Only for a random detector ($f=50\%$), the 
shedding coefficient is significantly reduced, as one cannot learn the 
correlation between data features and rumours.

\subsection{End-to-end Evaluation}
\label{sec:exp_ablation}

Finally, we report on an experiment that targets an end-to-end evaluation of 
our framework, under a latency threshold of $\theta = 1s$, averaged over all 
datasets. \autoref{tab:exp_ablation} presents the result for four ablation 
settings, in which we identify the importance of our strategies for \emph{load 
shedding} and \emph{rumour detector}, respectively. 
In the first setting, we 
run our full approach with the proposed strategies for load shedding and rumour 
detection, achieving an accuracy value of $0.91$ under a guaranteed latency 
value of $1s$. The second setup comprises 
the \emph{static} strategy for rumour detection without any shedding. This 
setup results in an accuracy value of $0.96$ and a latency of $6720s$, which 
violates the threshold. 
The third setup includes the 
\emph{dynamic} strategy for rumour detection with \emph{random shedding}. Here, 
the accuracy is $0.79$, under a guaranteed latency of $1s$. The last setting 
combines our shedding strategy with the \emph{static} rumour detector, which 
yields an accuracy of $0.86$, under a latency 
of $3617s$.

\begin{table}[!h]
\scriptsize
\centering
\caption{Ablation tests.}
\label{tab:exp_ablation}
\vspace{-7pt}
\scalebox{0.85}{
\begin{tabular}{cccc}
\toprule
Shedding Strategy & Rumour Detector & $F_{\beta}$-score (Prec$|$Rec) & latency (s) \\
\midrule
\textbf{Our} & \textbf{Dynamic} & \textbf{0.91} (0.92$|$0.90) & \textbf{$\sim$1} \\
None & Static & 0.96 (0.95$|$0.97) & 6720 \\
Random & Dynamic & 0.79 (0.84$|$0.74) & $\sim$1 \\
Our & Static & 0.86 (0.84$|$0.89) & 3617 \\
\bottomrule
\end{tabular}
}
\end{table}

In general, a good shedding strategy helps to achieve timely rumour detection,  
without sacrificing too much accuracy. Moreover, our proposed strategy for 
streaming rumour detection also improves the accuracy for streaming data. 
In \autoref{tab:exp_ablation}, we also report the respective precision 
and recall values for reference.

\section{Related Work}
\label{sec:related}

%

\sstitle{Multi-modal social graph}
The content of social networks can be modelled by a multi-modal social graph 
aka heterogeneous 
information networks~\cite{shi2017survey,fang2020survey,nguyen2021structural}. Some models capture
real-world entities, such as users and posts,
while others represent derived data elements, such as
topics and communities~\cite{li2017finding,toan2019deep,hung2017computing,inproceedings}. Existing work on
rumour detection in social networks focuses on propagation patterns of known
phenomena~\cite{friggeri2014rumor,zubiaga2018detection} or known
events~\cite{zhao2015enquiring}. This setting is orthogonal to our work,
since we strive for the detection of rumour, a phenomenon that emerges on 
social networks but is not known a priori~\cite{trung2020adaptive,nguyen2020entity,ren2022prototype}. 
Moreover, designing an algorithm that provides a latency guarantee for 
large-scale graphs is challenging~\cite{sahu2019ubiquity,das2019incremental,nguyen2014reconciling}, 
but is particularly necessary for applications in social 
networks~\cite{cong2018minimizing,trung2020comparative,nguyen2013batc,zheng2020pm,dong2019multiple}.

\sstitle{Rumour Detection}
Most means for rumour detection in social networks, surveyed 
in~\cite{zubiaga2018detection,shu2017fake,shu2019detecting}, are not suited for 
real-time applications. They rely on large training datasets and hence neglect 
evolving characteristics of rumours. Moreover, most traditional detection 
features emerge only after rumours affected many users. To overcome these 
limitations, methods for streaming rumour detection 
have been 
proposed~\cite{shu2019fakenewstracker,srijith2017sub,wang2017early,ding2019interspot},
 with most of them being based on pattern 
matching~\cite{wang2015bigdetecting,zhao2015enquiring,liu2015multi}. Yet, these 
approaches cannot give latency guarantees and may detect rumours solely with a 
significant delay~\cite{to2018survey,nguyen2020factcatch}. 
Our approach works under latency constraints and sheds unimportant data that do 
not contribute to rumour detection, regardless of the specific technique for 
rumour detection. \edit{Moreover, we change the nature of pattern matching from offline to online to make the detection even more efficient.}

\sstitle{Anomaly scoring}
Anomaly scoring techniques can be
divided into two categories: point-based or group-based~\cite{yu2016survey,zellag2014consistency,thang2015evaluation}. Point-based
methods work on individual data items to identify the signals that separate 
anomalies from the rest of the data~\cite{ihler2006adaptive,nguyen2019maximal,toan2018diversifying}.
Group-based techniques, in turn, identify subsets of individuals that 
collectively have different characteristics compared to any other 
subset~\cite{chen2014non,yu2015glad,muandet2013one,hung2019handling}.
Anomaly scoring is a reasonable strategy for rumour detection since rumours 
often propagate collectively over social networks. For example, 
\cite{chen2014non} addresses a similar use case, but neglects anomalies 
related to feature differences between entities. Most similar to our work 
is~\cite{tam2019anomaly}, which identifies subgraphs of social networks that 
are deemed to be rumours. However, the approach requires memory that grows in 
the size of the data to process, which is intractable in a streaming setting. 
\edit{Our work proposes a streaming anomaly computation via sketch structures while still guaranteeing a bounded quality.}

%
%

\sstitle{Load shedding}
In streaming processing, load shedding helps to cope with bursty input 
rates~\cite{oluwasuji2018algorithms,lee2018parallel}. While many approaches focus on queries 
for relational and sequential data stream 
processing~\cite{slo2019espice,HeBN14,ZhaoHW20}, load shedding is also useful 
for social networks that generate enormous volumes of high-velocity streaming 
data~\cite{zhao2021eires}. Our work is the first to propose a load shedding approach for rumour 
detection to provide latency guarantees. 
\edit{In particular, we design a coefficient modeling for each data element that is closely coupled with the proposed streaming pattern matching via occurrence utility.}

%

\section{Conclusion}
\label{sec:conclusion}

In this paper, we proposed an approach for streaming rumour detection for 
social networks under latency constraints. In the presence of high-velocity 
data streams, we argue for best-effort processing: Our goal is to detect the 
majority of rumours quickly. To this end, we took existing ideas on 
anomaly-based rumour detection, which identify local and global anomalies for 
propagation structures as captured by rumour patterns, as a starting point. 
Specifically, we lifted 
these ideas from a static setting to a streaming setting. \edit{We presented an 
algorithm for matching graph-based patterns in asymptotically constant time via a pattern index, 
along with a model for streaming anomaly scoring of individual entities and 
whole 
subgraphs via sketch structures}. \edit{Moreover, our load shedding goes beyond the state-of-the-art by developing a streaming model to 
capture the correlation between the coefficient of streaming data and rumours.} 
Based thereon, we proposed a \emph{coefficient-based} load shedding strategy to 
drop low-coefficient data when the system exceeds the latency threshold.
Experiments on large-scale real-world data showed that our approach is more 
effective and more efficient than baseline strategies. Our approach also turned 
out to be robust over diverse application settings~\cite{quoc2013evaluation,nguyen2014pay,nguyen2015result,
hung2013leveraging,nguyen2017argument,nguyen2020monitoring}.

\end{document}